\documentclass{aa}  

\usepackage{graphicx,natbib}
\usepackage{txfonts}
\begin{document}

   \title{Anisotropy in the all-sky distribution of galaxy morphological types}

   \author{Behnam Javanmardi
          \inst{1,2}\fnmsep\thanks{Member of the International Max Planck Research School (IMPRS) for Astronomy and Astrophysics at the Universities of Bonn and Cologne.}\fnmsep\thanks{\email{behnam@astro.uni-bonn.de}}
\and
          Pavel Kroupa\inst{2,3}
          }

   \institute{Argelander Institut für Astronomie der Universität Bonn, Auf dem Hügel 71, Bonn, D-53121, Germany
   \and
	     Helmholtz-Institut fuer Strahlen und Kernphysik, Nussallee 14-16, Bonn, D-53115, Germany
	     \and
	      Charles University in Prague, Faculty of Mathematics and Physics, Astronomical Institute, V  Hole\v{s}ovi\v{c}k\'ach 2, CZ-180 00 Praha 8, Czech Republic
             }

   \date{}

 
   \abstract{We present the first study of the isotropy of the all-sky distribution of morphological types of galaxies in the Local Universe out to around 200 Mpc using more than 60,000 galaxies from the HyperLeda database. We use a hemispherical comparison method in which by dividing the sky into two opposite hemispheres, the abundance distribution of the morphological types, $T$, are compared using the Kolmogorov-Smirnov (KS) test and by pointing the axis of symmetry of the hemisphere pairs to different directions in the sky, the KS statistic as a function of sky coordinates is obtained. For three samples of galaxies within around 100, 150, and 200 Mpc, we find a significant hemispherical asymmetry with a vanishingly small chance of occurring in an isotropic distribution. Astonishingly, regardless of this extreme significance, the observed hemispherical asymmetry for the three distance ranges is aligned with the Celestial Equator at the $97.1\%-99.8\%$ and with the Ecliptic at the $94.6\%-97.6\%$ confidence levels, estimated using a Monte Carlo analysis. Shifting $T$ values randomly within their uncertainties has a negligible effect on this result. When a magnitude limit of $B\leq 15$ mag is applied to the above mentioned samples, the galaxies within 100 Mpc show no significant anisotropy after randomization of $T$. However, the direction of the asymmetry in the samples within 150 and 200 Mpc and the same magnitude limit is found to be within an angular separation of 32 degrees from $(l,b)=(123.7, 24.6)$ with 97.2\% and 99.9\% confidence levels, respectively. This direction is only 2.6 degrees away from the Celestial North Pole. Unless the Local Universe has a significant anisotropic distribution of galaxy morphologies aligned with the orientation or the orbit of the Earth (which would be a challenge for the Cosmological Principle), our results show that there seems to be a systematic bias in the classification of galaxy morphological types between the data from the Northern and the Southern Equatorial sky. Further studies are absolutely needed to find out the exact source of this anisotropy.}

   \keywords{ Galaxies: statistics, Galaxies: structure, Galaxies: abundances, Astronomical data bases, Methods: statistical, Methods: data analysis}

  \maketitle
%

\section{Introduction}
Galaxies appear in various shapes and are observed to have a range of different properties and one of the main ways of studying their evolution is to classify them based on those observed features. The most widely known classification of galaxies, which is famous as the ``Hubble's tuning fork''\footnote{Based on a classification originally published in \citet{reynolds} and later by \citet{hubble26}, and on the tuning fork of \citet{jeans}. Its famous form was later presented in \citet{hubble}. See \citet{block} for a historical note.}, categorizes the (mostly nearby) galaxies into a range of morphological types based on bulge/disk domination. This classification was later revised by \citet{devauc59} who added a numerical value to each Hubble type and also to the intermediate stages. 

The morphology of galaxies is closely linked to their physical properties and those of their environments \citep{sandage75,kormendy82,bergh98,abraham98,calvi12} and is one of the important tools for studying galaxy formation and evolution. Bulge formation scenarios depend on galaxy formation models \citep{hopkins2010,kroupa15,corredoira16,combes16}, distinct galaxy types are observed to have very different stellar populations and star formation rates \citep{grebel11} and different spectral properties \citep{sanchez}, and their inner structure like bar and bulge types is connected with their observed kinematics \citep{molaeinezhad}. For a study on the evolution of the Hubble sequence see \citet{serrano} and for a recent review on galaxy morphology see \citet{buta13}.

Though the majority of the bright nearby galaxies fit in the Hubble's tuning fork, the high redshift galaxies detected by deep surveys and the low surface brightness dwarf galaxies whose number is increasing with various surveys inside and outside the Local Group (e.g. by \citeauthor{des} \citeyear{des} and \citeauthor{javanmardi16} \citeyear{javanmardi16}, respectively), are hard to be classified using the standard morphological classification system \citep{naim97,abraham01}.   

One of the most widely used classification schemes is that of \citeauthor{devauc59} compiled in the Third Reference Catalogue of Bright Galaxies (RC3) \citep{devauc91} from which many other catalogs extract the morphological types of different galaxies. As pointed out by \citet{makarov14}, visual inspection has been the main method of classification in RC3.\\

The need for automated morphology classification has been known since decades \citep{naim95} and has been exercised in recent years (see e.g. the recent studies by \citeauthor{company} \citeyear{company} and \citeauthor{kuminski16} \citeyear{kuminski16}, and references there in). \citet{nair10} and \citet{baillard11} used the data from the Sloan Digital Sky Survey (SDSS) and attempted to improve the visual classification of galaxy morphologies with the aim of paving the way for automated galaxy classification by providing training sets and calibration samples \citep[for other catalogs of galaxy morphologies see][and \citeauthor{psychogyios} \citeyear{psychogyios}]{fukugita07,shibuya,herrera,krywult, poudel}. In general, the goal is to achieve a catalog of the morphological types of the observed galaxies as complete and systematic-free as possible and to have a well-defined classification method applicable to future large galaxy surveys. Such a catalog is obviously crucial for studies of galaxy formation and evolution.\\

In this work, and for the first time, we search for possible deviations from isotropy in the all-sky distribution of the morphological types of galaxies within around 200 Mpc using the HyperLeda database. Based on the Cosmological Principle (generally understood to be confirmed by most of the observations so far), on sufficiently large scales the properties of the Universe, including the distribution of galaxy types, should be statistically isotropic. Therefore, deviations from isotropy can be a hint of systematic issues in the morphological classification of galaxies or in the homogenization of catalogs. \\

On the other hand, it is vital to re-inspect the assumption of isotropy with various observations \citep{maartens11} and this is one of the motivations of our study. If a significant deviation from cosmic isotropy is detected and confirmed by various data sets, cosmology will face a major paradigm change. During the last decade, probing isotropy in all-sky extragalactic data has become a vibrant research topic and continues to deliver interesting results. \cite{tegmark}, \citet{eriksen} and \citet{hansen04} reported some large scale anisotropies (hemispherical asymmetry and quadrupole-octopole alignment) in the Cosmic Microwave Background (CMB) radiation data from the \textit{WMAP} satellite. These CMB ``anomalies'' were recently confirmed by the \citet{planck} suggesting that they are not artifacts caused by the detectors or data-reduction procedures \citep[see also][and \citeauthor{mukherjee} \citeyear{mukherjee}]{akrami,rassat,copi15}. \citet{javanmardi15} found an anisotropy in the magnitude-redshift relation of high redshift Type Ia Supernovae (SNe Ia) that is significantly aligned with the direction of the CMB dipole and very close to the CMB quadrupole-octopole alignment \citep[for similar studies see][and references there in]{carvalho15,bengaly16,migkas,lin}. Also an inconsistency between the amplitude of the observed dipole in the distribution of radio galaxies and the value expected from the CMB dipole was reported in \citet{singal}, \citet{rubart} and \citet{tiwari}. For a recent review on various isotropy studies see \citet{zhao}. \\

The isotropy of the spatial distribution of galaxies has been probed by various authors \citep{gibelyou,yoon,appleby,alonso,bengaly16b}. These studies have found some mild anisotropies with different directions but none of them reported a significant deviation.

In our analysis, we consider three distance ranges separately; galaxies with radial velocity less than 7,000, 10,000 and 14,000 km/s (equivalent to around 100, 150 and 200 Mpc from us, respectively\footnote{Assuming the Hubble constant value of $H_0=70.0$ km s$^{-1}$ Mpc$^{-1}$.}). Based on the standard model of cosmology, at such distance scales and specially beyond $\approx$150 Mpc \citep{marinoni12}, the distribution of galaxies should be statistically isotropic. For each distance range, we separate the galaxies by dividing the sky into two opposite hemispheres and compare their morphological type distribution using the Kolmogorov-Smirnov test. By pointing the axis of symmetry of our hemispheric cut towards different directions on the sky and repeating the test, we find the pair of hemispheres with the largest difference in the distribution of morphological types and quantify the significance of the difference.\\

The rest of this paper is organized as follows. In Section \ref{sec:data} we describe our sample from the HyperLeda database. Our method of analysis is explained in Section \ref{sec:method}. We present the results in Section \ref{sec:results}, discuss them critically in Section \ref{sec:dis} and finally we summarize and conclude in Section \ref{sec:conc}.

\section{Data}\label{sec:data}
HyperLeda \citep{paturel03,paturel03b,makarov14} is a large database of more than three million objects, around 1.5 million of which are confirmed galaxies that uniformly cover the entire sky except from the region around the Galactic plane. This database contains different observed properties like magnitude, surface brightness, color, redshift, and morphological type of galaxies. These measurements have been homogenized to standard systems and are updated when new measurements become available \citep{makarov14}.\\

The morphological types, $T$, in the HyperLeda are homogenized values based on compilation of morphology measurements from the literature. These types are in the de Vaucouleurs scale \citep{devauc91} ranging from $T=-5$ for elliptical galaxies to $T=10$ for irregular galaxies. Most of these measurements are obtained by visual inspection of optical images. After the homogenization (which also takes into account the quality of each measurement), the best possible value and its uncertainty for each galaxy is determined. They are not limited to integer values and have an accuracy up to the first decimal digit in the HyperLeda database (e.g. $T=2.2\pm0.7$ for the galaxy NGC3368). The information on the instruments used for these measurements are not kept in the database and one needs to refer to the individual works for that\footnote{The authors obtained some of the information in this paragraph by a private communication with Dmitry Makarov (one of the team members of the HyperLeda) in August 2016.}.\\

For the purpose of this study, we limit our sample to all the confirmed galaxies that have a radial velocity in the CMB reference frame, $V_{CMB}$, smaller than 14,000 km/s. This velocity corresponds to a distance of about 200 Mpc. This gives us around 200,000 galaxies from the HyperLeda database. For each galaxy, we obtain its Equatorial and Galactic coordinates, its morphological type (if available), its corresponding uncertainty, $\sigma_T$, corrected apparent total B magnitude, and absolute B-band magnitude, M$_B$, from the database. \\

\begin{figure}
 \begin{center}
  \includegraphics[scale=0.36]{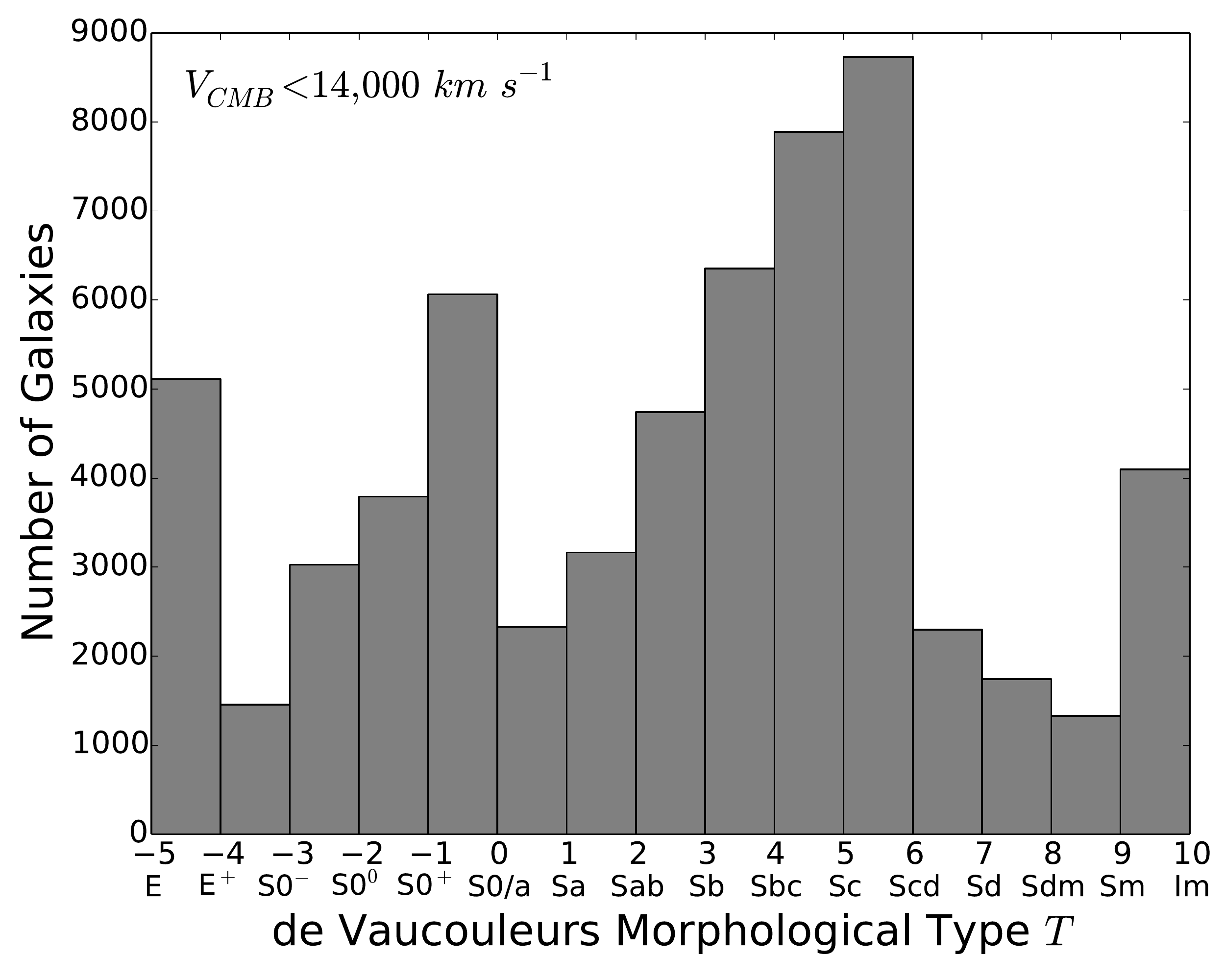}
  \caption{The number distribution of the $T$ values for the galaxies in the HyperLeda database with $V_{CMB}<14,000$ km/s and a measured M$_B$. In all the bins only the left edge is included in the counting (e.g. for the the first bin $-5\leq T < 4$) except for the last bin which also includes the right edge (i.e. $9\leq T \leq 10$). This is the same for the rest of the histogram plots in this paper. \label{fig:main_hist}}
 \end{center}
\end{figure}

\begin{figure*}
 \begin{center}
  \includegraphics[width=\textwidth]{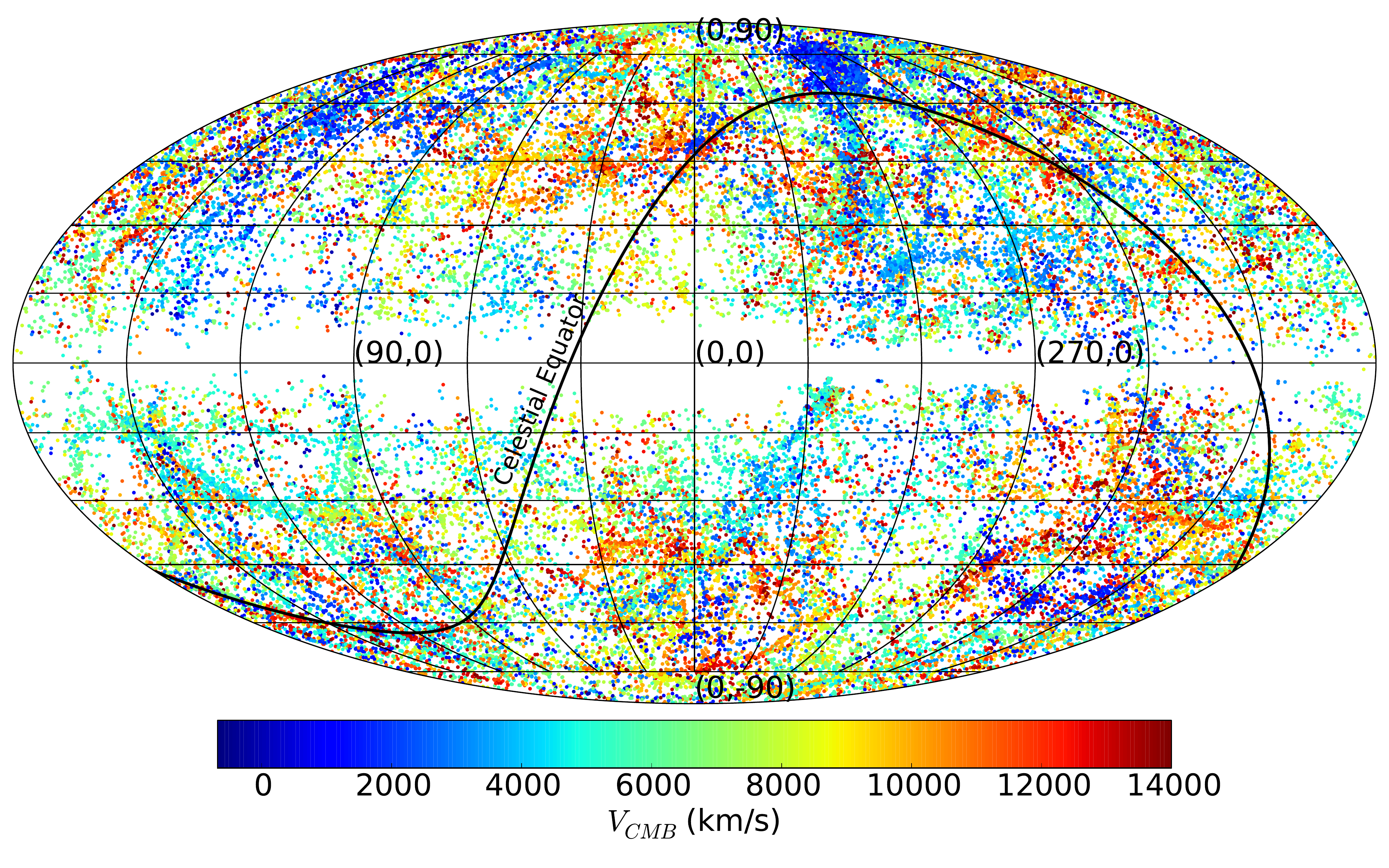}
  \caption{The sky distribution of all the galaxies in the HyperLeda database with $V_{CMB}<14,000$ km/s that have $T$ and M$_B$, in the Galactic coordinate system. The color code is radial velocity, $V_{CMB}$, in km/s and the Celestial Equator is shown by a solid black curve. The sample uniformly covers all the sky (except from the regions in and around the disk of our Galaxy).\label{fig:sky_dist_v}}
 \end{center}
\end{figure*}

More than 65,000 of these galaxies ($\approx$33 \% of the sample) have a measured morphological type assigned to them in the database\footnote{Our sample was downloaded from the database in May 2016. The database might have been updated since then.}. In order to increase the robustness of our analysis, we exclude the $\approx$3000 galaxies (around 5\% of the galaxies with $T$) for which an absolute B-band magnitude is not available\footnote{We will consider the effect of including them in our analysis in Section \ref{sec:include_Mb}.}. This condition gives us our final sample of more than 62,000 galaxies which we use for our statistical analysis. We refer to it as \textit{the whole sample} hereafter. The faintest object in the sample has $B=23.5$ mag, but only around 2500 galaxies in the sample have $B\geq17.0$ mag\footnote{We will also apply magnitude limits to our sample in Section \ref{sec:mag_limit}.}. The number distribution of the $T$ values of the whole sample is shown in Figure \ref{fig:main_hist} and its sky distribution in the Galactic coordinate system is shown in Figure \ref{fig:sky_dist_v} with the color code being $V_{CMB}$ in units of km/s. We see that galaxies are distributed uniformly across the sky (except for the region around the Galactic plane) and we see no trace of any particular survey. 

\section{Method}\label{sec:method}

\subsection{Hemispherical comparison} \label{sec:hc}
Let us consider a particular hemisphere of the sky whose pole is pointing at a direction $\hat{r}$ with longitude and latitude $(l,b)$ in the Galactic coordinate system. We then separate the galaxies into two groups; those that are inside that hemisphere, and the ones that are in the opposite hemisphere whose pole is pointing towards $-\hat{r}$ with coordinates $(l+180^{\circ},-b)$. We then compare the morphological distribution of these two sets of galaxies. The aim is to vary $\hat{r}$ to point towards different directions on the sky and repeat the hemispheric division and distribution comparison in order to find the pair of hemispheres with the largest difference in the distribution of $T$ and to calculate the significance of the difference. We start from the northern Galactic hemisphere, so $-\hat{r}$ will be in the southern hemisphere. We pixelise the sky using a HEALPix\footnote{Hierarchical Equal Area isoLatitude Pixelation, http://healpix.sourceforge.net/} grid \citep{gorski} with 768 pixels (or directions). This gives us 384 pairs of opposite hemispheres for our analysis.

\subsection{Kolmogorov-Smirnov test}
A simple but powerful statistical method for quantifying the level of consistency of the distribution of two data sets is the Kolmogorov-Smirnov (KS) test. This non-parametric method is used to test if two data sets come from the same parent distribution. After separating the galaxies into two groups using the hemispheric cut (explained in Section \ref{sec:hc}), we obtain the cumulative distribution functions, $S_{\hat{r}}(T)$ and $S_{-\hat{r}}(T)$, of the morphological types in the hemispheres pointing at $\hat{r}$ and $-\hat{r}$, respectively. These distribution functions are normalized so that their largest value is equal to unity. The KS statistic as a measure of the difference between these two distributions is simply the maximum value of the absolute difference between $S_{\hat{r}}(T)$ and $S_{-\hat{r}}(T)$ \citep{press}:

\begin{equation}\label{eq:dks}
 D(\hat{r})=\underset{-5\leq T \leq10}{max} |S_{\hat{r}}(T)-S_{-\hat{r}}(T)|
\end{equation}

We find $D(\hat{r})$ for all the 384 hemisphere pairs to determine the direction with the largest $D$.

\begin{figure*}
 \begin{center}
  \includegraphics[scale=0.42]{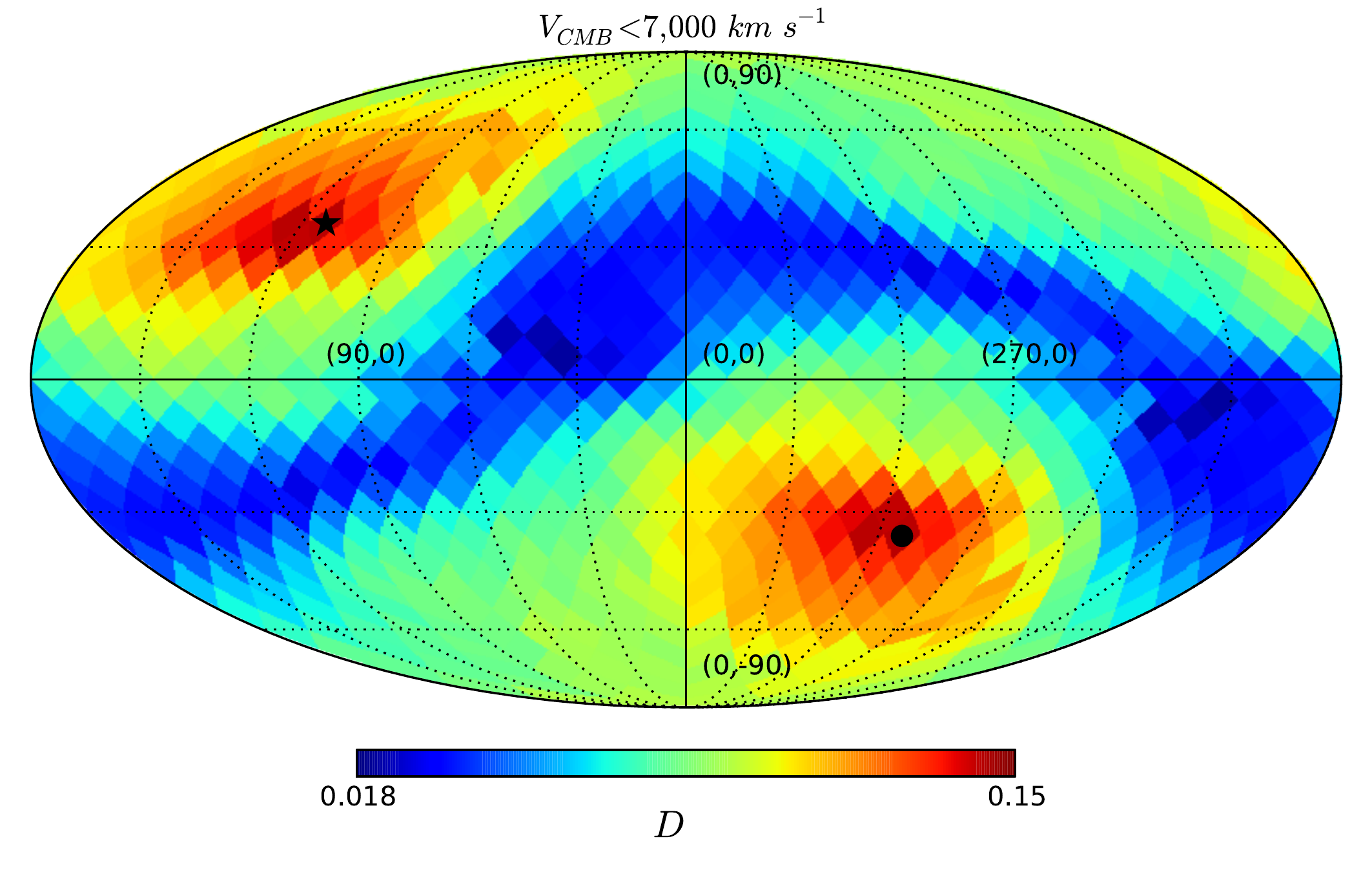}
  \includegraphics[scale=0.42]{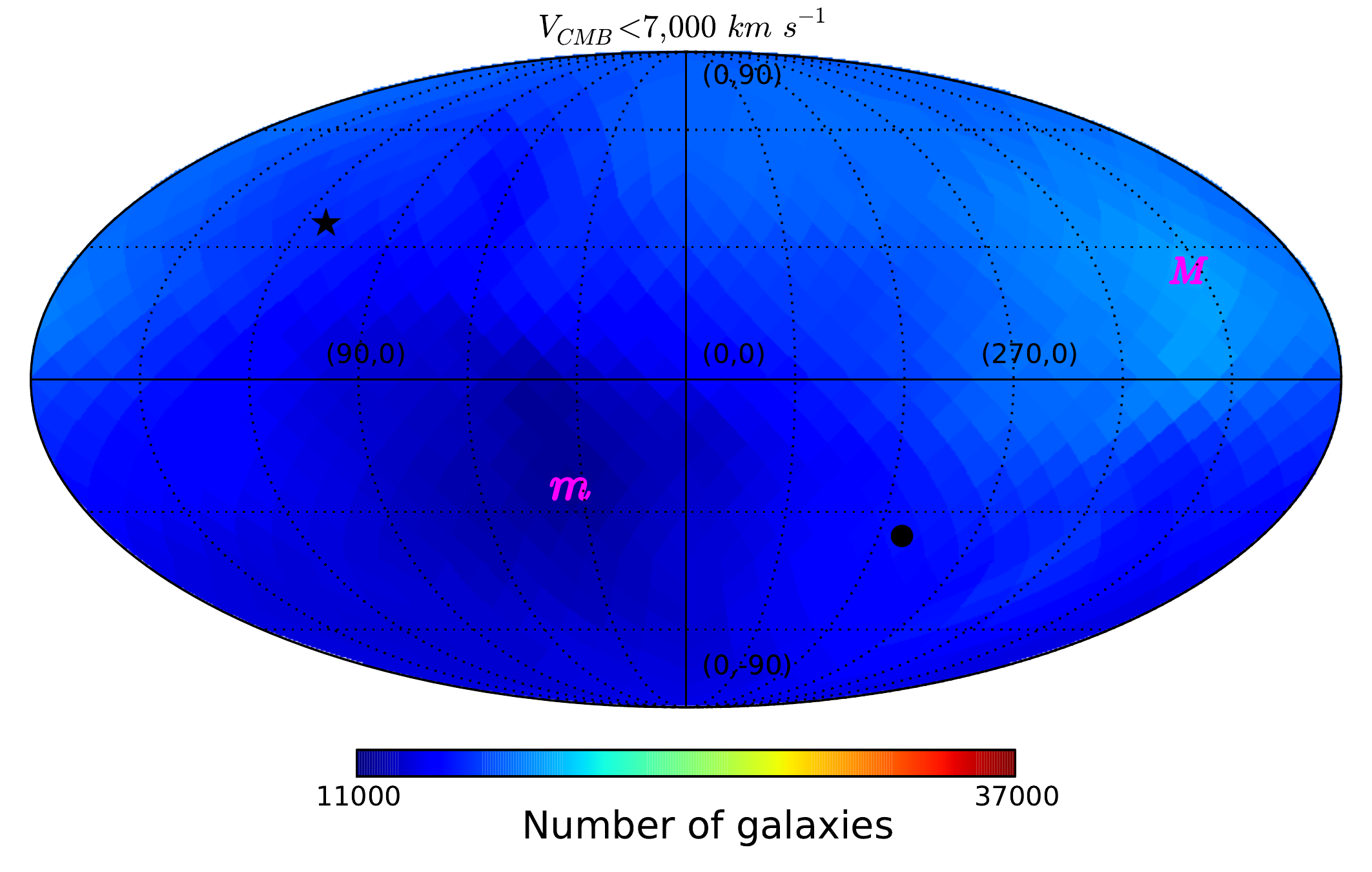}
    \includegraphics[scale=0.42]{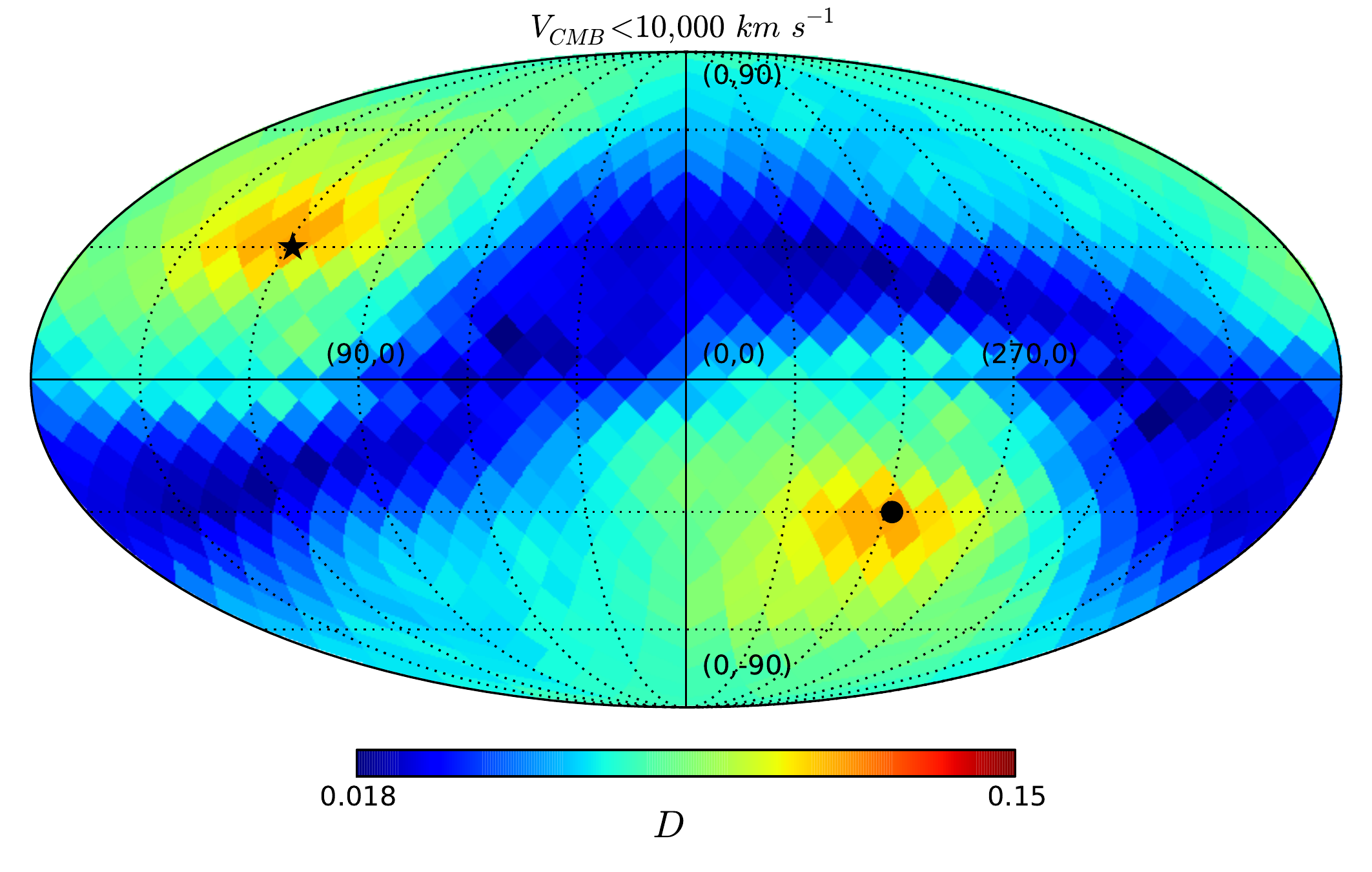}
    \includegraphics[scale=0.42]{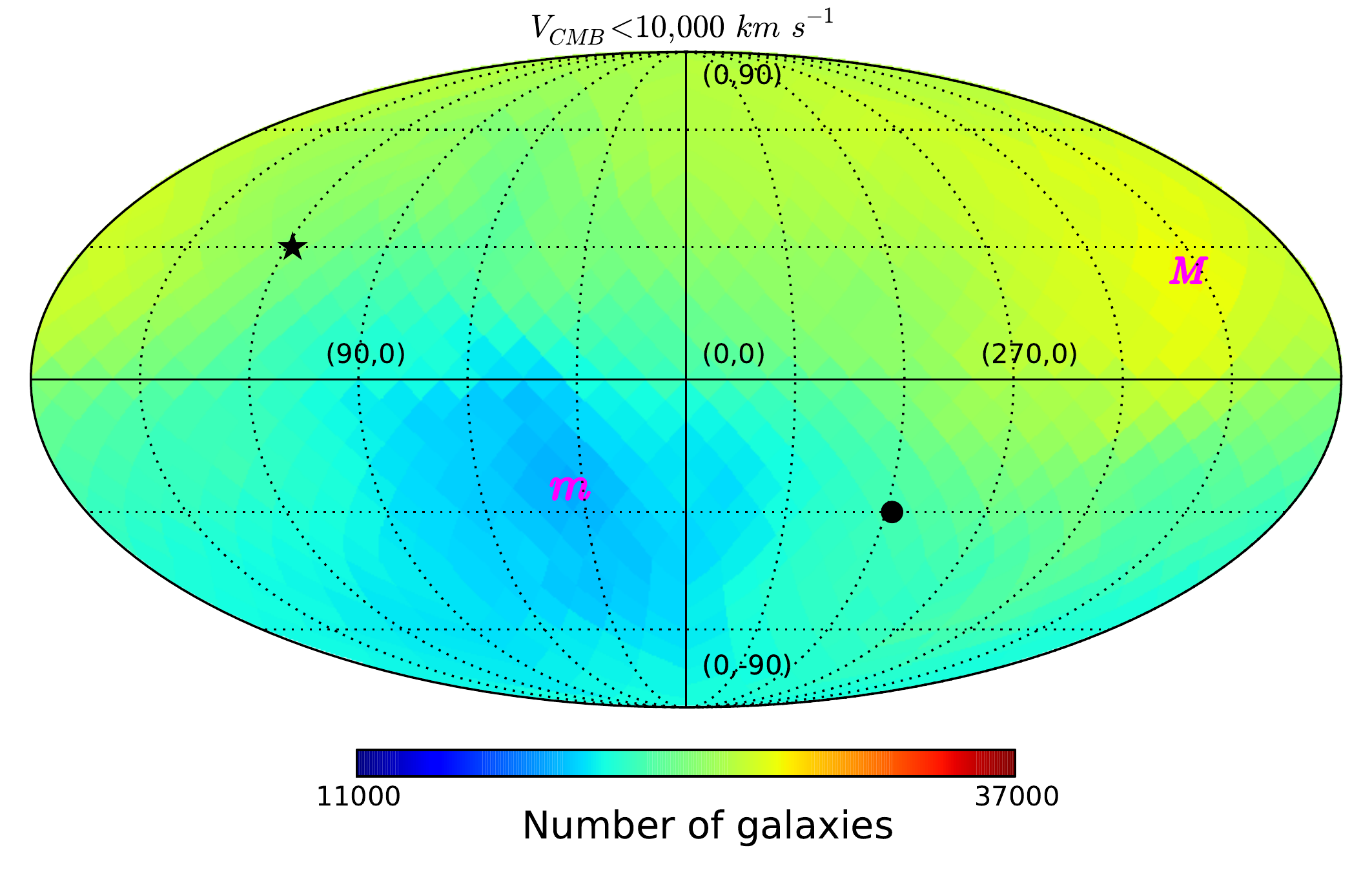}
  \includegraphics[scale=0.42]{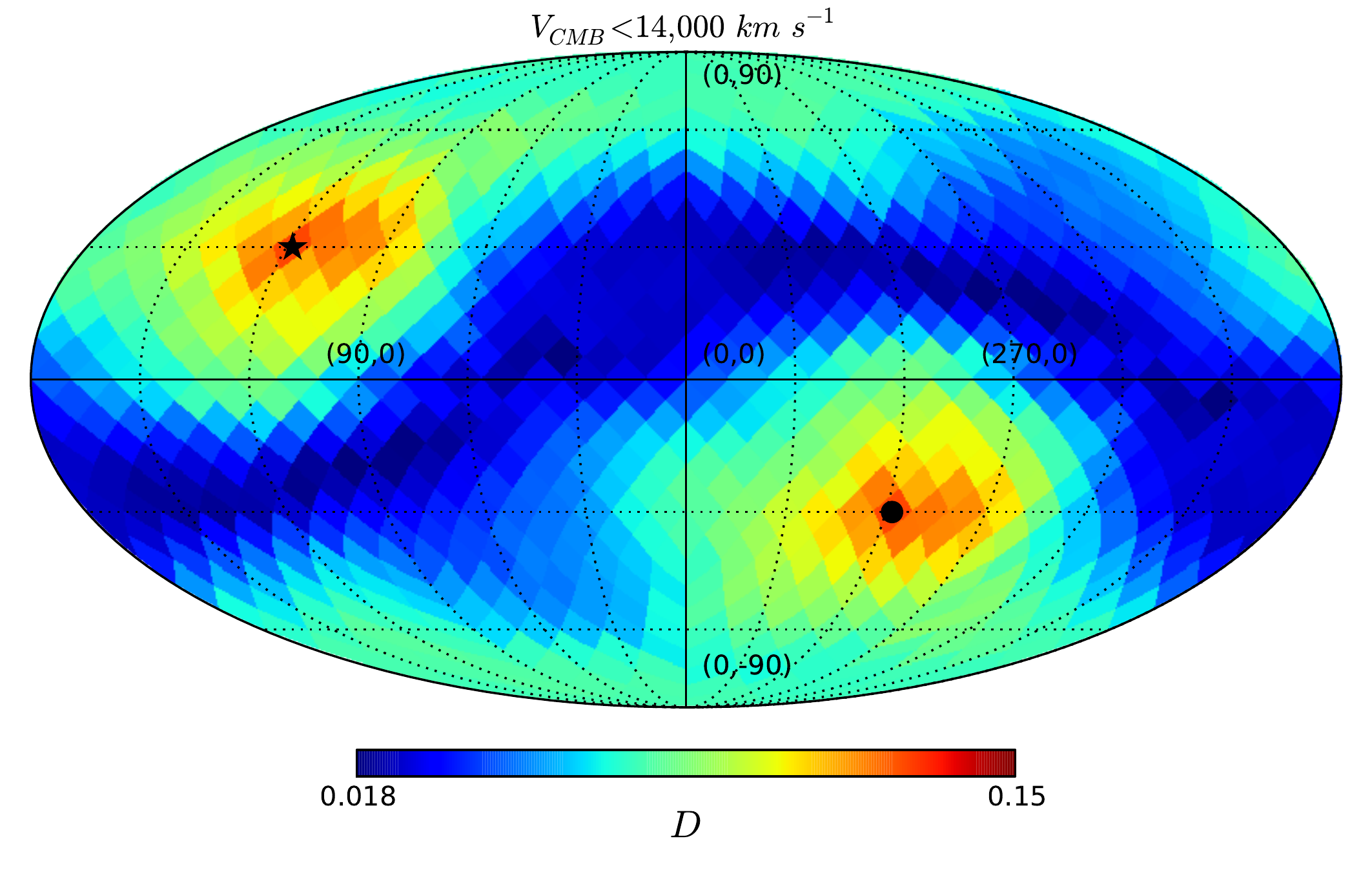}
  \includegraphics[scale=0.42]{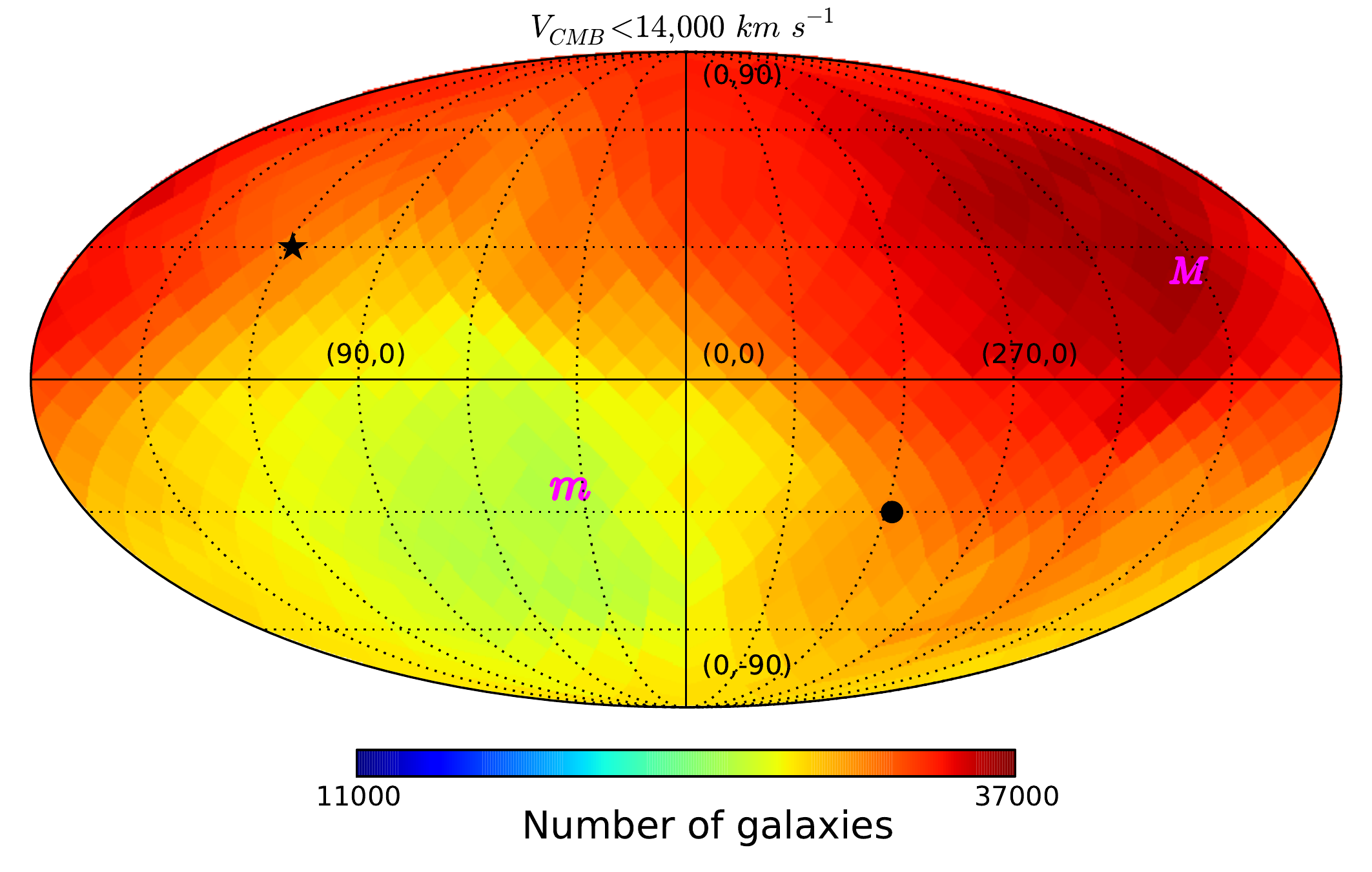}

  \caption{Left: The sky maps of $D$ from the KS test in the Galactic coordinate system from top to bottom for $V_{CMB}<$7,000, 10,000 and 14,000 km/s, respectively. The value in each pixel is obtained from the KS comparison of the pair of hemispheres with axis of symmetry along the center of that pixel. Obviously, the opposite direction of each pixel has exactly the same value. On each map, the direction $\hat{r}_{max}$ is marked by a star and its opposite direction, $-\hat{r}_{max}$, by a circle. Right: The variation of the number of galaxies of each sample across the sky also in the Galactic coordinate system. The value in each pixel is the number of galaxies in a hemisphere whose pole is pointing towards that pixel. The directions with the minimum and the maximum number are denoted by small $m$ and capital $M$, respectively. We see that the directions with the largest difference in the distribution of morphologies are very different from those with the largest difference in the number of galaxies (see Section \ref{subsec:hc}).\label{fig:KSD_maps}}
  
 \end{center}
\end{figure*}

\section{Results}\label{sec:results}
\subsection{Hemispherical comparison and the directions with the largest difference} \label{subsec:hc}
Sky maps of $D(\hat{r})$ for the three radial velocity ranges are shown in the left panel of Figure \ref{fig:KSD_maps}. The value in each pixel is obtained by applying the KS test on the hemisphere whose pole is pointing towards the center of that pixel and its opposite hemisphere. Obviously the $D$ for each direction $\hat{r}$ and its opposite direction $-\hat{r}$ is exactly the same since it is a measure of the difference between them. On each map, the direction with the largest difference, $\hat{r}_{max}$, and its opposite direction, $-\hat{r}_{max}$, are marked by a star and a circle, respectively. It can be seen that the general trend and the $\hat{r}_{max}$ in all the three maps are very close to each other. 

In Table \ref{t:results}, we list the coordinates of the $\hat{r}_{max}$ and the corresponding KS results, $D_{max}$, for the three $V_{CMB}$ ranges. The number of total galaxies in the HyperLeda database within each radial velocity range, $N_{tot}$, and the number of galaxies that have both $T$ and $M_B$ and were used in our analysis, $N_a$, are also listed in this Table. We see that $D_{max}$ decreases when increasing the distance from $V_{CMB}<$7,000 to 10,000 km/s, but increases again when going to $V_{CMB}<$14,000 km/s, while $\hat{r}_{max}$ remains unchanged in the latter step. 

In this table, we also list the number of galaxies in the pair of hemispheres corresponding to $D_{max}$, namely $N_{\hat{r}_{max}}$ and $N_{-\hat{r}_{max}}$. We see that their difference is small for all the three distance ranges, but to check if these directions are related or are close to the directions with the largest difference in the number of galaxies, we can look at the variation of the number of galaxies in each sample across the sky. In the right panel of Figure \ref{fig:KSD_maps}, the number of galaxies in the hemispheres whose poles are pointing towards each pixel is plotted. We also show $\hat{r}_{max}$ and $-\hat{r}_{max}$ as in the left panel. The directions with the minimum and the maximum number of galaxies are marked with small $m$ and capital $M$, respectively. Although these are only the galaxies that have a $T$ value, the change in the number is smooth; the maximum, $M$, is consistently close to the direction that the Local Group moves towards \citep[with respect to the CMB rest frame, see e.g.][]{gibelyou}, and the minimum, $m$, is in the opposite direction. Hence, the directions corresponding to $D_{max}$ are very different from those with the largest difference in the number of galaxies.

\begin{figure*}
 \begin{center}
  \includegraphics[width=\textwidth]{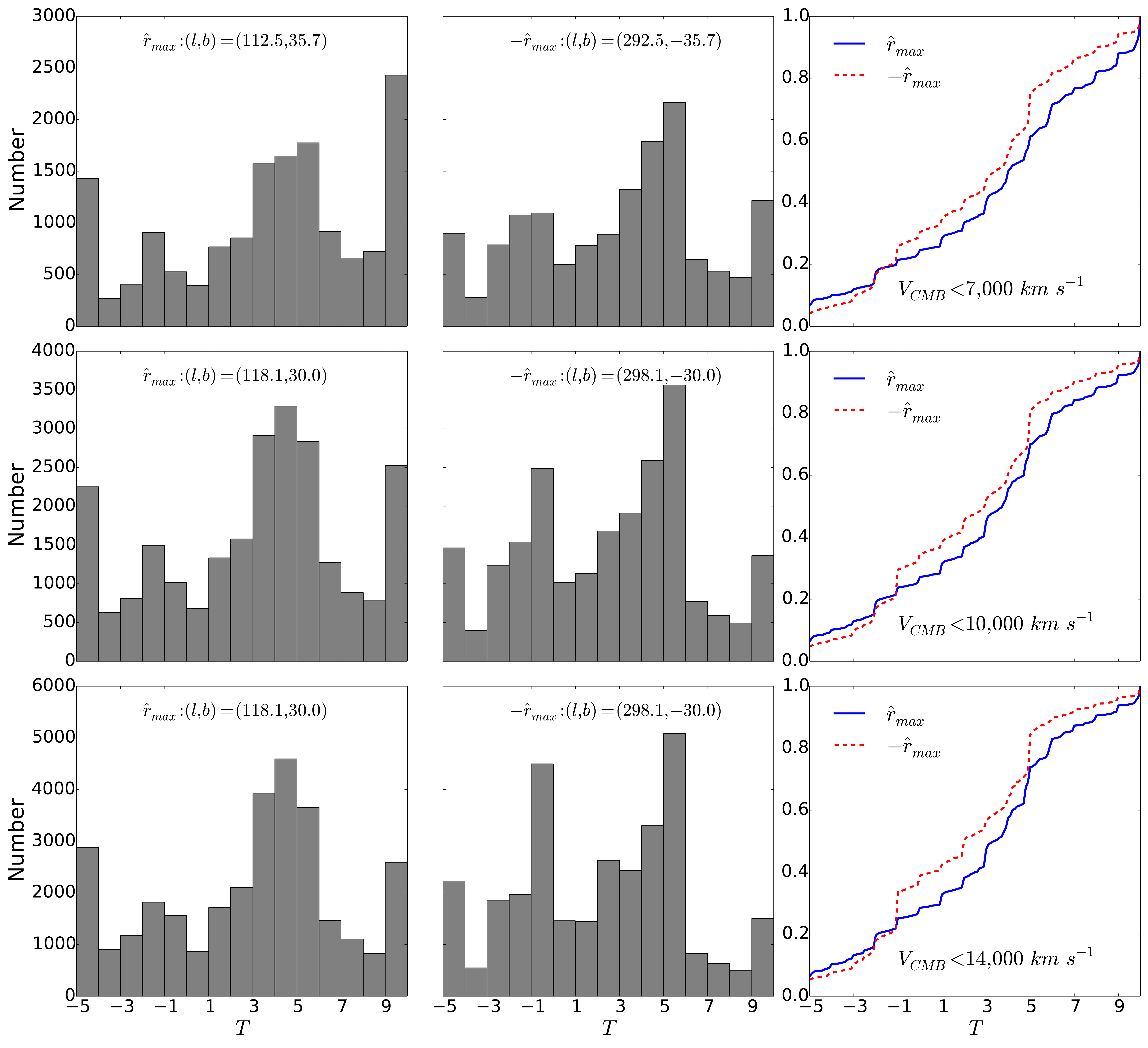}
  \caption{The distribution of morphologies in the hemisphere pairs with the largest difference, i.e. towards $\hat{r}_{max}$ and $-\hat{r}_{max}$, in the first and second columns and their corresponding cumulative distributions in the third column from top to bottom for $V_{CMB}<$7,000, 10,000 and 14,000 km/s, respectively. The prominent differences are at $9\leq T \leq 10$ and  $-1\leq T < 0$. \label{fig:hists_and_cumuls}}
  
 \end{center}
\end{figure*}

In the first two columns of Figure \ref{fig:hists_and_cumuls} we show the distribution of $T$ in the hemispheres towards $\hat{r}_{max}$ and $-\hat{r}_{max}$ for the three $V_{CMB}$ ranges. Their cumulative distributions are shown in the third column of this figure. We can see two prominent differences between the morphological distribution of galaxies in these opposite hemispheres. First, the number of galaxies with $9\leq T \leq 10$ (i.e. Sm and Im types) in the direction of $\hat{r}_{max}$ is around twice as large as the number of these types in the direction $-\hat{r}_{max}$ in all the three $V_{CMB}$ ranges. Second, the number of galaxies with $-1\leq T < 0$ (i.e. S0$^+$ types) in the direction of $-\hat{r}_{max}$ is more than twice the number of these galaxies in the direction of $\hat{r}_{max}$, again for all the three $V_{CMB}$ ranges. In the latter case, a clear jump in the cumulative distribution functions is visible. Another increase in this function is at $T=5$ (i.e. Sc galaxies) and their number is larger towards $-\hat{r}_{max}$ in all the three $V_{CMB}$ ranges. 

Finally, we emphasize that the results are almost the same in all the three distance ranges.

\begin{table*}
 \begin{center}
  \caption{Our three radial velocity ranges and their corresponding estimated distances, the total number of galaxies from the HyperLeda database, $N_{tot}$, for each range and the number of galaxies $N_{a}$ used in our analysis (i.e. having $T$ and M$_B$). The Galactic coordinates of the directions, $\hat{r}_{max}$, having the largest difference and their corresponding KS statistic, $D_{max}$, are listed in columns 5 and 6, respectively. Also the number of galaxies in the pair of hemispheres with the largest difference, $N_{\hat{r}_{max}}$ and $N_{-\hat{r}_{max}}$ are listed. Note that these two numbers are similar for all the three cases.\label{t:results}}
  
  \begin{tabular}{lccccccc}
  $V_{CMB}/(\rm{km/s})<$ & $\approx d$ (Mpc) & $N_{tot}$ & $N_{a}$ & $\hat{r}_{max}=(l,b)$ & $D_{max}$ & $N_{\hat{r}_{max}}$ & $N_{-\hat{r}_{max}}$\\
  \hline
 7,000 & 100 & 62,314 & 29,830 &(112.5, 35.7) & 0.143 & 15,268 & 14,562 \\
 10,000 & 150 & 118,858 & 46,533 & (118.1, 30.0) & 0.115 & 24,307 & 22,226 \\
 14,000 & 200 & 198,971 & 62,125 & (118.1, 30.0) & 0.128 & 31,194 & 30,931 \\

  \hline
  \end{tabular}
 \end{center}
\end{table*}

\subsection{Significance of the observed asymmetry}\label{subsec:significance}
Given the assumption that the distribution of galaxies should be statistically isotropic in the distance ranges under consideration and therefore that the distribution of morphologies should be statistically similar in different directions, we calculate the significance of the observed anisotropy using the KS distribution and different Monte Carlo (MC) analyses. 

\subsubsection{The KS distribution}\label{sec:ks_pvalue}
The null hypotheses here is that the morphological types in the opposite hemispheres towards $\hat{r}_{max}$ and $-\hat{r}_{max}$ come from the same parent distribution. The probability that this hypotheses is true is equal to the probability of obtaining a difference larger than (or equal to) the observed difference; $p(D_{max})=Prob_{KS}(D\geq D_{max})$. This can be calculated using the standard two-sided Kolmogorov-Smirnov distribution which is a function of the number of data points \citep[see][for the details]{press}. The third column of Table \ref{t:pvalues}, contains the probabilities, $p(D_{max})$, obtained for each $D_{max}$ (which are listed again in this table to ease the comparison). These values are dramatically small and basically consistent with zero. For a better comprehension of these values, we can compare them to those for the directions with the smallest difference, $D_{min}$, which are also listed in Table \ref{t:pvalues} as $p(D_{min})$. In other words, $p(D_{min})$ corresponds to the pair of opposite hemispheres with the most similar distributions of $T$. We see that $p(D_{min})$ is many orders of magnitude larger than $p(D_{max})$ for all the three distance ranges.\\

\subsubsection{Monte Carlo: Isotropic realizations}\label{sec:shuffling}
In an effort to investigate the significance calculated from the KS distribution even further, we can use MC analyses. In the first method, we create mock galaxy samples with an isotropic distribution of morphologies by shuffling the position of the galaxies in the analysis while keeping the sky distribution fixed to the original one. Although shuffling can (in some cases) give conditions like having a late type galaxy at the centre of a galaxy cluster (where usually an elliptical sits), it is still a good way to have an estimate of the distribution of $D$ in the case of a fully random sky distribution. We create 1000 random realizations with an isotropic distribution and repeat the hemispherical comparison method and find the pair of hemispheres with the largest $D$ for each of the realizations. For all the three distance ranges, the number of realizations having a $D_{max}$ equal to or larger than the observed values is zero. Actually, the largest $D_{max}$ values obtained in 1000 realizations are 0.029, 0.027 and 0.019 for $V_{CMB}<$7,000, 10,000 and 14,000 km/s, respectively, which are much smaller than the observed values of $D_{max}$.\\

\subsubsection{Monte Carlo: Random sampling}
In the second MC approach, we randomly draw $N_{\hat{r}_{max}}$ galaxies from each sample (regardless of their coordinates) and compare their $T$ distribution with the rest of the galaxies in the sample (whose number amounts to $N_{-\hat{r}_{max}}$) with the KS test. We repeat this 100,000 times and again in none of the random samples we find a $D$ as large as or larger than the observed $D_{max}$ values. In this case, the largest $D$ values obtained in 100,000 realizations are 0.028, 0.022 and 0.020 for $V_{CMB}<$7,000, 10,000 and 14,000 km/s, respectively, which are similar to those obtained by the shuffling method. \\

These MC results (and the ones from Section \ref{sec:shuffling}) were expected from the calculated probabilities from the KS distribution and together they show that the observed values of $D_{max}$ are extremely improbable to occur out of an isotropic distribution of galaxy morphologies.

\begin{table}
 \begin{center}
  \caption{The probability $p(D_{max})$ (from the KS distribution) that the observed difference, $D_{max}$, occurs if the morphologies in the pair of hemispheres are from the same parent distribution. For comparison, we list the smallest value obtained from the hemispherical comparison, $D_{min}$, and its corresponding probability, $p(D_{min})$, for each radial velocity range (see Section \ref{sec:ks_pvalue}). Note that $p(D_{min})$ is orders of magnitude larger than $p(D_{max})$ for all the three cases.  \label{t:pvalues}}
  
  \begin{tabular}{lcccc}
  $V_{CMB}/(\rm{km/s})$< & $D_{max}$ & $p(D_{max})$ & $D_{min}$ & $p(D_{min})$\\
  \hline
 7,000 &  0.143 & $10^{-133}$ & 0.022 & 1.7$\times 10^{-3}$\\
 10,000 &  0.115 & $10^{-134}$ & 0.018 & 8.2$\times 10^{-4}$ \\
 14,000 &  0.128 & $10^{-221}$ & 0.018 & 6.7$\times 10^{-5}$ \\

  \hline
  \end{tabular}
 \end{center}
\end{table}

\begin{figure*}
 \begin{center}
  \includegraphics[width=\textwidth]{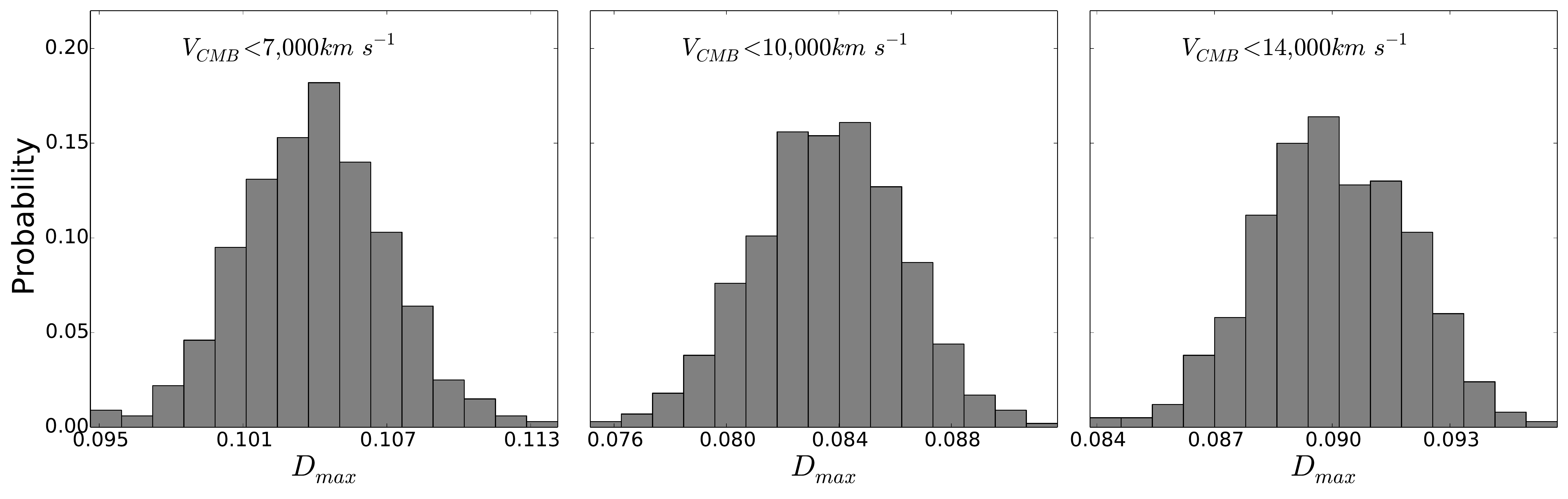}
  \caption{The probability distribution of $D_{max}$ for 1000 realizations in which $T_i$ of each galaxy is replaced by $T_i+\Delta T$ where $\Delta T$ is randomly drawn from a Gaussian distribution with zero mean and standard deviation $\sigma_{T_{i}}$. From left to right for $V_{CMB}<$7,000, 10,000 and 14,000 km/s, respectively. See section \ref{sec:sigma_t}. \label{fig:sigma_T}}
  
 \end{center}
\end{figure*}

\subsection{Alignment with the Celestial Equator and the Ecliptic}
The observed significant anisotropy has a peculiar feature that asks for even further analysis. The direction $\hat{r}_{max}$ for all the three distance ranges is very close to the Celestial North Pole (CNP) and also relatively close to the Ecliptic North Pole (ENP), or in other words, the plane separating the hemispheres corresponding to $D_{max}$, is aligned with the Celestial Equator and the Ecliptic. The angular separations between $\hat{r}_{max}$ and the CNP, $\alpha_{CNP}$, and the ENP, $\alpha_{ENP}$, are listed in Table \ref{t:alignment}. Such an alignment is totally unexpected even if the result from the KS test would not have been significant. Since $\hat{r}_{max}$ is close to both the CNP and the ENP, we quantify the significance of the alignment to both of them. 

\begin{table}
 \begin{center}
  \caption{The angular separation between $\hat{r}_{max}$ and the Celestial, $\alpha_{CNP}$, and the Ecliptic, $\alpha_{ENP}$, North Poles. The fraction of 1000 random realizations in which $\hat{r}_{max}$ is as aligned as or more aligned than the observed direction with the CNP, $f_{CNP}$, and the ENP, $f_{ENP}$, is also listed.    \label{t:alignment}}
  
  \begin{tabular}{lcccc}
  $V_{CMB}/(\rm{km/s})$< & $\alpha_{CNP}$ & $\alpha_{ENP}$ & $f_{CNP}$ &  $f_{ENP}$\\
  \hline
 7,000 &  12$^{\circ}$.3 & 14$^{\circ}$.7 & 0.029 &  0.024\\
 10,000 &  5$^{\circ}$.1 & 18$^{\circ}$.8 & 0.003 &  0.054 \\
 14,000 &  5$^{\circ}$.1 & 18$^{\circ}$.8 & 0.002 &  0.052 \\

  \hline
  \end{tabular}
 \end{center}
\end{table}

\subsubsection{Significance of the alignment from Monte Carlo analysis}
Using the 1000 isotropic realizations from the MC analysis in Section \ref{sec:shuffling}, we find the fractions, $f_{CNP}$ and $f_{ENP}$, of isotropic samples in which the direction with the largest $D$ has the same or better alignment with the CNP and the ENP, respectively. These values are also listed in Table \ref{t:alignment}. For obtaining these fractions we do not consider the significance of the anisotropy (i.e. the value of $D_{max}$) and we only take the direction into account. The fraction of random samples with better alignment and larger $D_{max}$ would be zero as is obvious from the results in Section \ref{subsec:significance}. Interestingly, $f_{CNP}$ shows that even regardless of the significance of the KS test, the hemispherical asymmetry is aligned with the Celestial Equator at the 97.1\%, 99.7\%, and 99.8\% confidence levels for $V_{CMB}<$7,000, 10,000 and 14,000 km/s, respectively. The alignment of the observed hemispherical asymmetry with the Ecliptic is at the $94.6\%-97.6\%$ confidence levels, depending on the distance range under considerations. Actually, the latter alignment is slightly more significant than the former one for the sample with $V_{CMB}<$7,000 km/s. It may be worth to note that one of the CMB anomalies, i.e. the hemispherical asymmetry in the power spectrum is also aligned with the Ecliptic plane \citep{planck}. \\

\subsection{Effect of shuffling $T$ within $\sigma_T$ on the significance and the direction of the anisotropy}\label{sec:sigma_t}
The uncertainty of the morphological types, $\sigma_T$, for the whole sample ranges between 0.1 and 6.6 with an average $\bar{\sigma}_T=1.9$. The majority of galaxies have $\sigma_T \approx 2.0$. When a $\sigma_T$ is available for a galaxy in the HyperLeda database, it means that several astronomers have classified that galaxy. All the galaxies in our sample have a $\sigma_T$ assigned to their $T$ values. By performing another MC analysis, we check the extent to which random uncertainties reduces the significance of the observed asymmetry and change its direction. To achieve this, we replace the morphological type $T_i$ of each galaxy by $T_i+\Delta T$ where $\Delta T$ is randomly drawn from a Gaussian distribution with zero mean and standard deviation $\sigma_{T_{i}}$. We then perform the hemispherical comparison on the new $T$ values. This process of randomization of $T$ and search for asymmetry is repeated 1000 times for the three distance ranges. Figure \ref{fig:sigma_T} shows the distribution of $D_{max}$ obtained from this MC analysis.\\

In 99.9\% of these realizations, $p(D_{max})$ is smaller than $1.5 \times 10^{-58}$, $4.2 \times 10^{-58}$, and $3.2 \times 10^{-96}$ for $V_{CMB}<$7,000, 10,000 and 14,000 km/s, respectively. For the $V_{CMB}<$7,000 km/s range, the direction $\hat{r}_{max}$ remains exactly the same in 73.4\% of the realizations while in the rest $\hat{r}_{max}$ is within $9^{\circ}.1$ of the observed direction. For the sample with $V_{CMB}<$10,000 km/s, 83.3\% of the cases yield the same $\hat{r}_{max}$ and the rest have an $\hat{r}_{max}$ within $7^{\circ}.4$ of the observed direction. And finally, for the distance range $V_{CMB}<$14,000 km/s, in all the 1000 realizations the direction $\hat{r}_{max}$ remains exactly the same as the observed direction. Therefore, shuffling $T$ within $\sigma_T$ has minor to no effect on the direction of the asymmetry found in these distance limited samples.\\

In Figure \ref{fig:sky_dist_t}, we show the distribution of galaxy morphologies for the three radial velocity ranges in our study in the Equatorial coordinate system\footnote{We note that all our analysis was done in the Galactic coordinate system, plotting the distribution of $T$ in the Equatorial coordinate system is only to help seeing the observed alignments easier.}. The black dotted line is the plane of the Ecliptic and the red solid line is the plane separating the pair of hemispheres corresponding to $D_{max}$ for each case. As can be seen, the observed hemispherical asymmetry is very close to both the Ecliptic plane and the Celestial Equator for $V_{CMB}<7,000$ km/s but is closer to the Celestial Equator for $V_{CMB}<10,000$ and 14,000 km/s. Although in the case of $V_{CMB}<14,000$ km/s the number of galaxies in the analysis increases by more than $30\%$ with respect to this number for $V_{CMB}<10,000$ km/s, the direction of $\hat{r}_{max}$ does not change. We can clearly see a region in the south of the Celestial Equator occupied mostly with early type galaxies and the northern sky is more populated by late type galaxies.

\section{Discussion}\label{sec:dis}
At the moment, our results cannot be compared with other studies because this is the first time that the isotropy of galaxy morphological types is investigated. The previous statistical studies of the morphological types \citep[e.g.][and \citeauthor{delap} \citeyear{delap}]{nair10} were limited to only a fraction of the sky. The studies on the isotropy of the number distribution of galaxies by \citet{gibelyou} and \citet{alonso}, and on the isotropy of the luminosity function of galaxies by \citet{appleby} did not report a significant deviation from isotropy, but their reported directions are relatively close to the direction of the CMB dipole. The studies by \citet{yoon} and \citet{bengaly16b} of the isotropy of the galaxy number counts, both found (with low significance) similar dipole directions of $(l,b)=(310, -15)$ and $(l,b)=(323, -5)$, that are different from the above mentioned studies, but are not far from the $-\hat{r}_{max}$ directions in our study. \\

\subsection{Sample selection effects}
Since the anisotropy we found in the distribution of morphological types is very significant and unexpectedly aligned with the CNP and ENP, it is necessary to scrutinize the results even further. In this section we discuss the cases regarding our sample that could affect the results. 

\subsubsection{Including the galaxies without $M_B$}\label{sec:include_Mb}
As we mentioned in Section \ref{sec:data}, the galaxies that do not have an available $M_B$ in the HyperLeda database, were excluded from our analysis. Although they constitute only about 5\% of the galaxies with $T$, we here check their effect on the results. We repeat the hemispherical comparison analysis for the whole sample of galaxies with $V_{CMB}<14,000$ km/s, this time including the $\approx 3000$ galaxies without $M_B$. The direction with the largest difference remains exactly the same and $D_{max}=0.124$. Hence, including these galaxies does not affect the final results. 

\subsubsection{Excluding galaxies with large error on $T$}
 Here we apply a strict cut and exclude all the galaxies with $\sigma_T \geq 3.0$ from the whole sample with $V_{CMB}<14,000$ km/s, this leaves us with 50,195 galaxies for the analysis. The hemispherical comparison again results in exactly the same direction and a slightly smaller difference, $D_{max}=0.121$. Therefore, excluding the galaxies with large error on $T$ does not have a significant effect on the results either.  

\subsubsection{Distance Tomography}
The subsamples we considered are not independent of each other, i.e. the whole sample with all the galaxies within $V_{CMB}<14,000$ km/s includes the other two samples with $V_{CMB}<$7,000 and 10,000 km/s. Here we repeat the hemispherical comparison analysis for the two independent subsamples of galaxies with 7,000$<V_{CMB}/(\rm{km/s})<$10,000 and 10,000$<V_{CMB}/(\rm{km/s})<$14,000. For the former sample, $D_{max}=0.132$ and $p(D_{max})=4.6\times 10^{-63}$, and for the latter sample, $D_{max}=0.203$ and $p(D_{max})=9.0\times 10^{-139}$. For both of them, the direction $\hat{r}_{max}$ is $(l,b)=(118.1, 30.0)$, i.e. exactly the same as those of the samples with $V_{CMB}<$10,000 and 14,000 km/s (see Table \ref{t:results}). The KS test shows that the distance range 10,000$<V_{CMB}/(\rm{km/s})<$14,000 has the largest asymmetry in the distribution of $T$. The overall results remain unchanged.
 
\begin{figure}
 \begin{center}
  \includegraphics[scale=0.22]{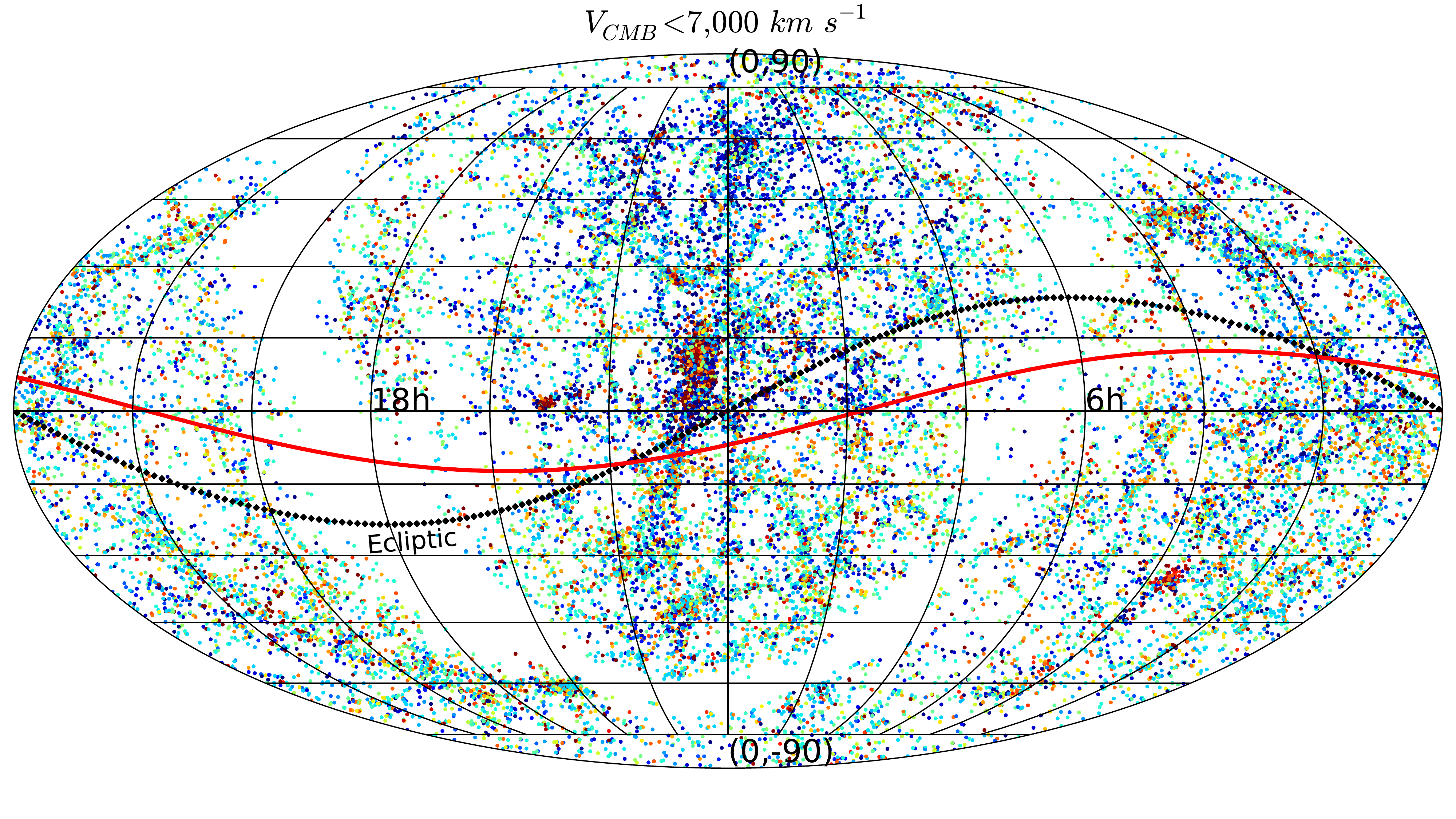}
  \includegraphics[scale=0.22]{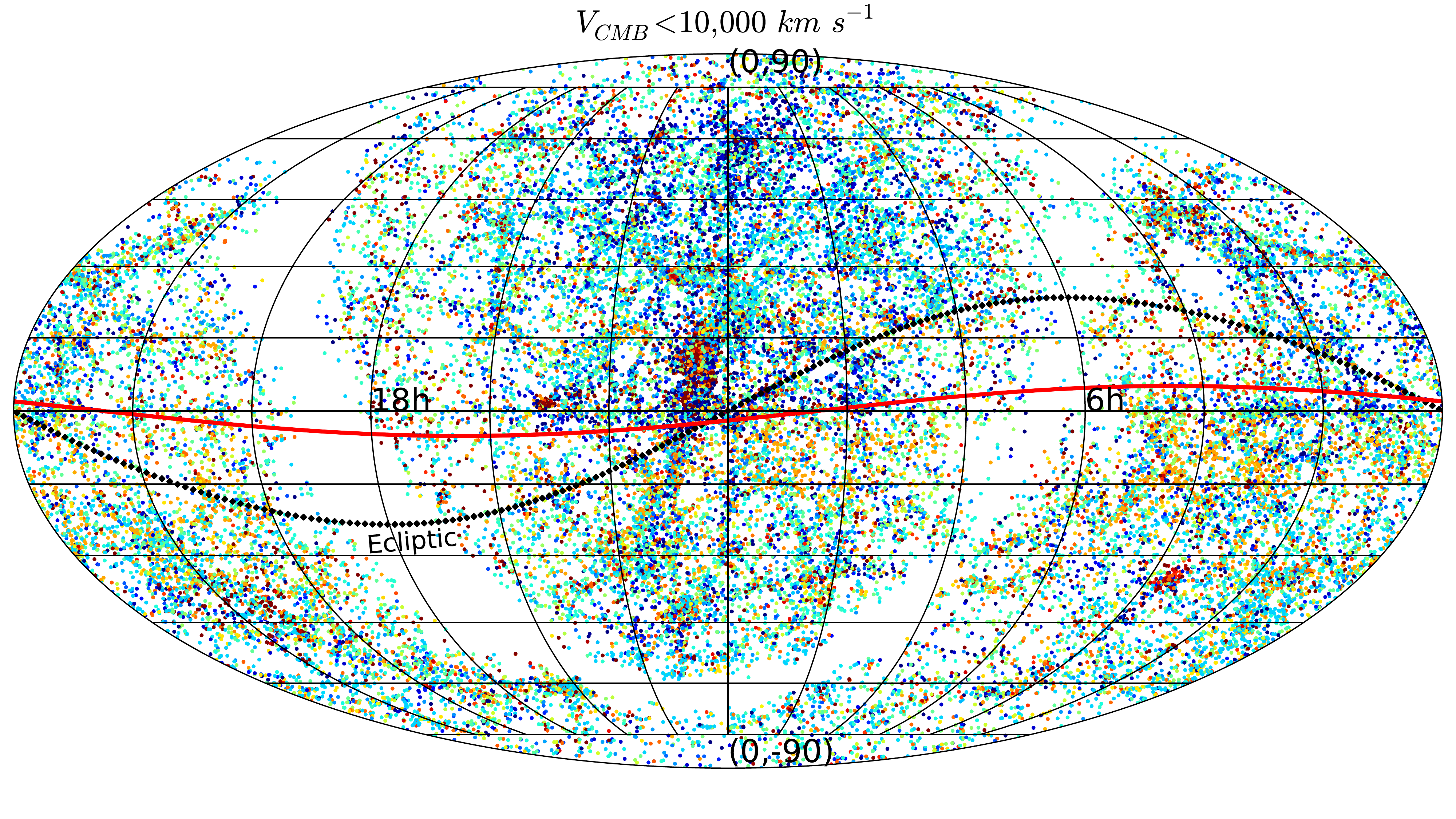}
    \includegraphics[scale=0.22]{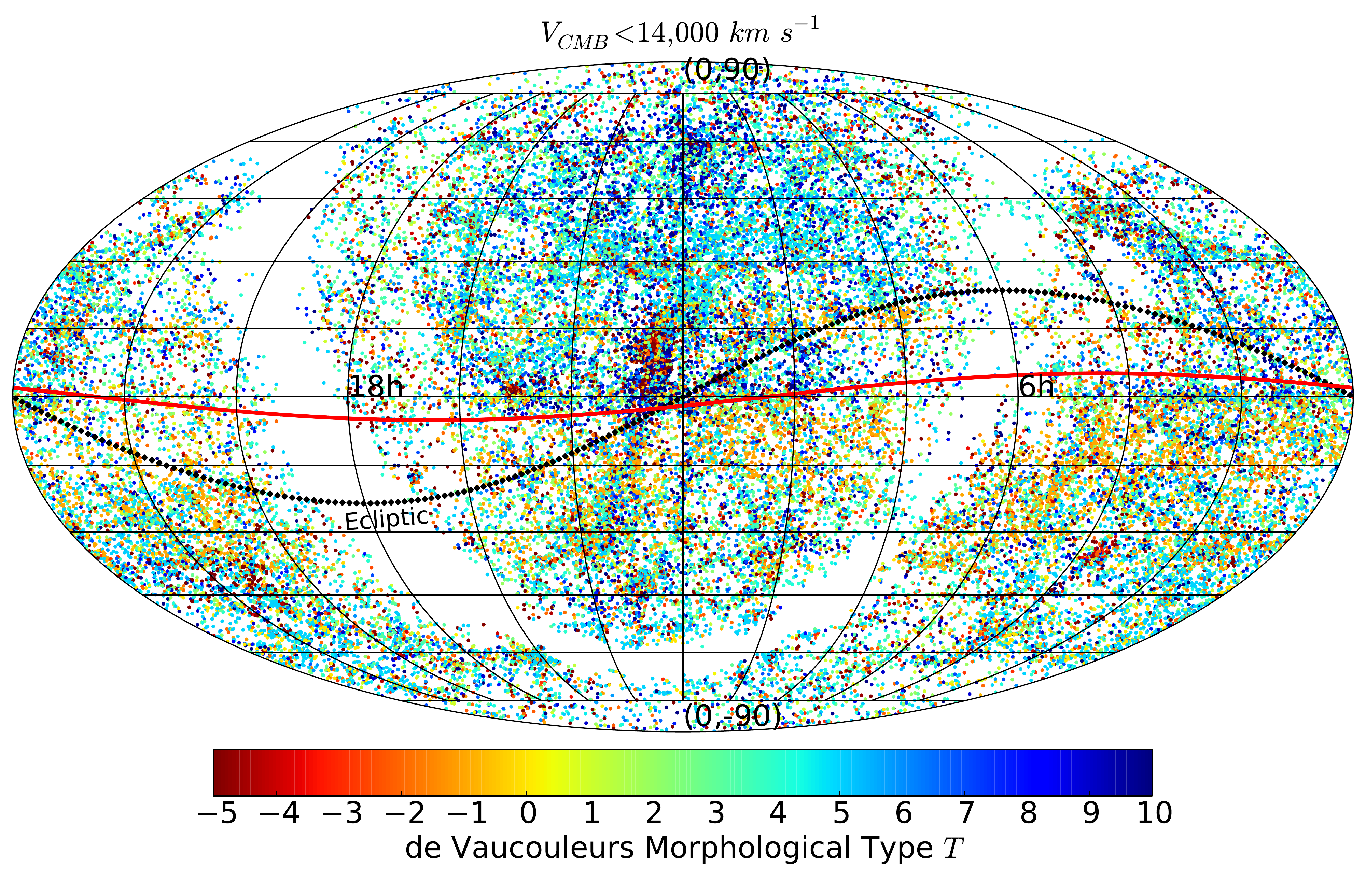}
  \caption{The Equatorial coordinate system distribution of all the galaxies in the HyperLeda database with measured $T$ and M$_B$, from top to bottom corresponding to $V_{CMB}<$7,000, 10,000 and 14,000 km/s, respectively. The color code is $T$, the empty parts are the region around the disk of the Milky Way, the plane of the Ecliptic is shown with a dotted line and the pair of hemispheres with the largest difference in the distribution of morphologies are divided by a red solid line.\label{fig:sky_dist_t}}
  
 \end{center}
\end{figure}
\subsubsection{Applying a magnitude limit to the sample}\label{sec:mag_limit}
The factor that is more probable to affect the results is that a large number of galaxies within 14,000 km/s in the HyperLeda database do not have a $T$ value. Had we known the morphological types of these galaxies, the result could be different. Unfortunately, this information is not available at the moment. \\

However, we can test the effect of completeness by limiting the analysis to all the galaxies brighter than a certain magnitude. We put a conservative magnitude limit and confine our hemispherical comparison analysis to the galaxies with B$\leq$ 15 mag. The galaxy sample from the HyperLeda with $V_{CMB}<14,000$ km/s and B$\leq$15 mag is around 99\% complete, i.e. around 99\% of galaxies in the database with these two conditions have a $T$ value. This is the same for the samples with $V_{CMB}<$7,000 and 10,000 km/s and the same magnitude limit. The results of the hemispherical comparison for the three distance ranges are shown in Table \ref{t:completeness}. We see that the $D_{max}$ values are smaller with respect to the values for the samples without the magnitude limit. Although the level of anisotropy decreases and the values of $p(D_{max})$ are orders of magnitude larger than the ones for the samples without the B$\leq$ 15 mag limit, they are still quite small and the KS result is still very significant (compare the values of $p(D_{max})$ to those of $p(D_{min})$ which are also listed in Table \ref{t:completeness}). The direction of $\hat{r}_{max}$ did not change for the range $V_{CMB}<$7,000 km/s, however for $V_{CMB}<$10,000 and 14,000 km/s, the $\hat{r}_{max}$ is farther away from the ENP, but interestingly is even more aligned with the CNP having only a $2^{\circ}.6$ angular separation. \\

We apply the MC method explained and used in Section \ref{sec:sigma_t} to check the effect of shuffling $T$ within $\sigma_T$ on the anisotropy observed in these samples. For $V_{CMB}<$10,000 and 14,000 km/s, 97.2\% and 99.9\% of the realizations have their $\hat{r}_{max}$ within 32$^{\circ}.0$ of the observed direction\footnote{We choose the value 32$^{\circ}.0$ because it is the maximum angular separation of $\hat{r}_{max}$ for 99.9\% of the realizations of the sample with $V_{CMB}<$14,000 km/s and B$\leq$ 15 mag.}, and all the realizations have their $p(D_{max})$ smaller than $2.2\times 10^{-5}$ and $6.8\times 10^{-7}$, respectively. However, in the case of the sample with $V_{CMB}<$7,000 km/s, only 17.0\% of the realizations have their $\hat{r}_{max}$ within 32$^{\circ}.0$ of the observed direction, the rest have their $\hat{r}_{max}$ towards various directions and ranging to more than 80$^{\circ}.0$ away from the observed $\hat{r}_{max}$. Although shuffling $T$ within $\sigma_T$ washes out the the asymmetry in the sample with $V_{CMB}<$7,000 km/s and B$\leq$ 15 mag, the anisotropy in the samples with $V_{CMB}<$10,000 and 14,000 km/s and B$\leq$ 15 (specially the latter), remain significantly close to the observed direction of $\hat{r}_{max}$.

\begin{figure*}
 \begin{center}
  \includegraphics[width=\textwidth]{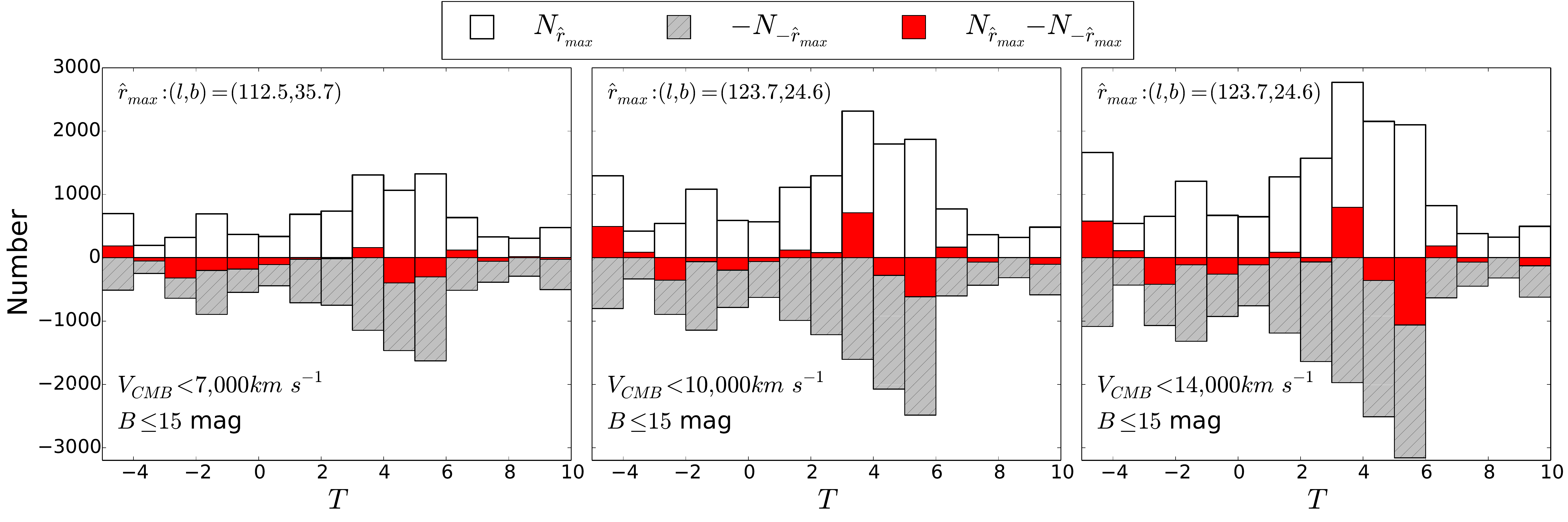}
  \caption{The distribution of morphologies in the hemispheres towards $\hat{r}_{max}$ (positive numbers) and $-\hat{r}_{max}$ (with negative numbers) for the samples with B$\leq$15 mag. From left to right corresponding to $V_{CMB}<$7,000, 10,000 and 14,000 km/s, respectively. The number difference between the two hemispheres for each $T$ is shown by red coloured bars. \label{fig:num_and_dif}}
  
 \end{center}
\end{figure*}

\begin{figure}
 \begin{center}
  \includegraphics[scale=0.22]{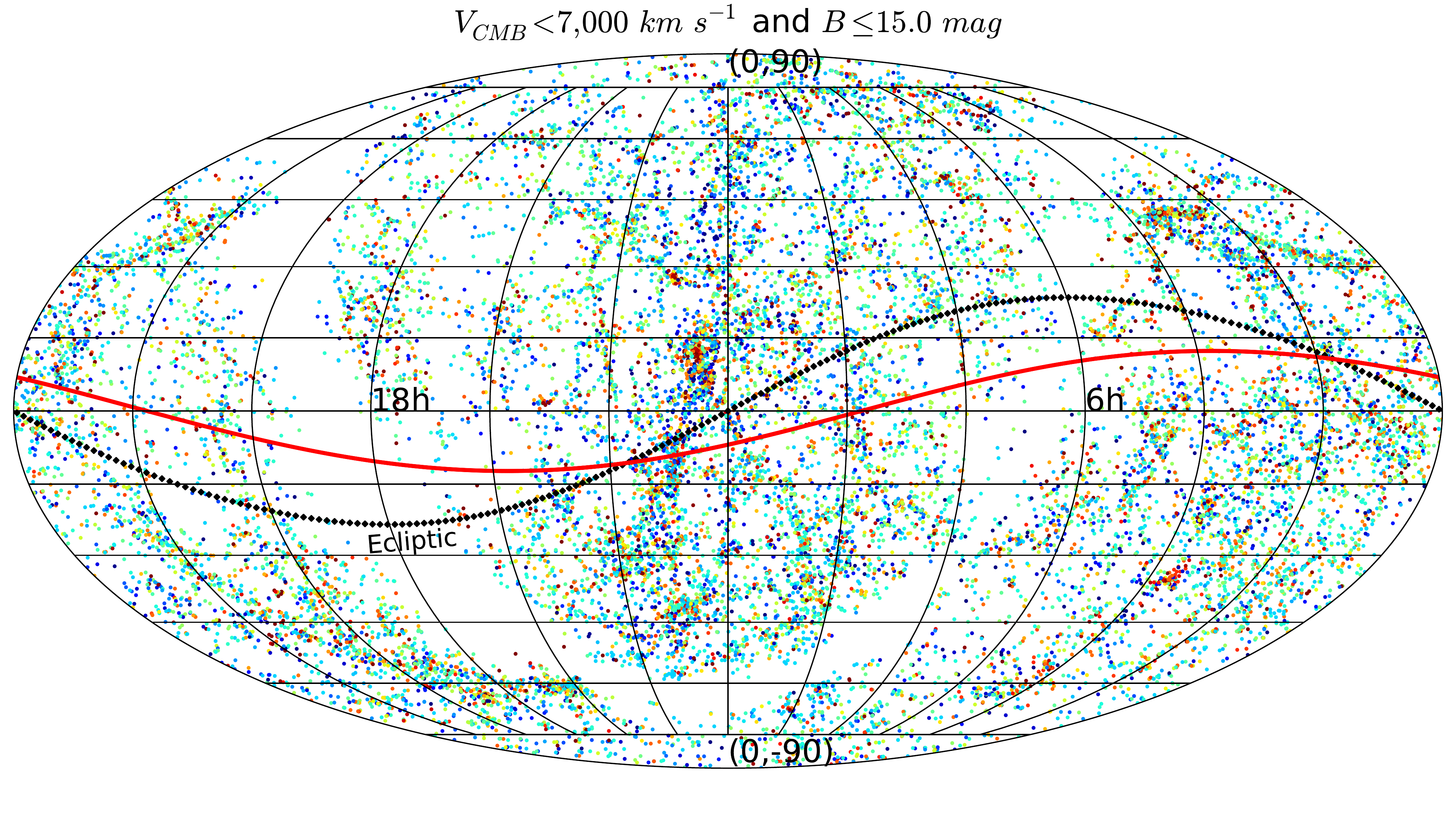}
  \includegraphics[scale=0.22]{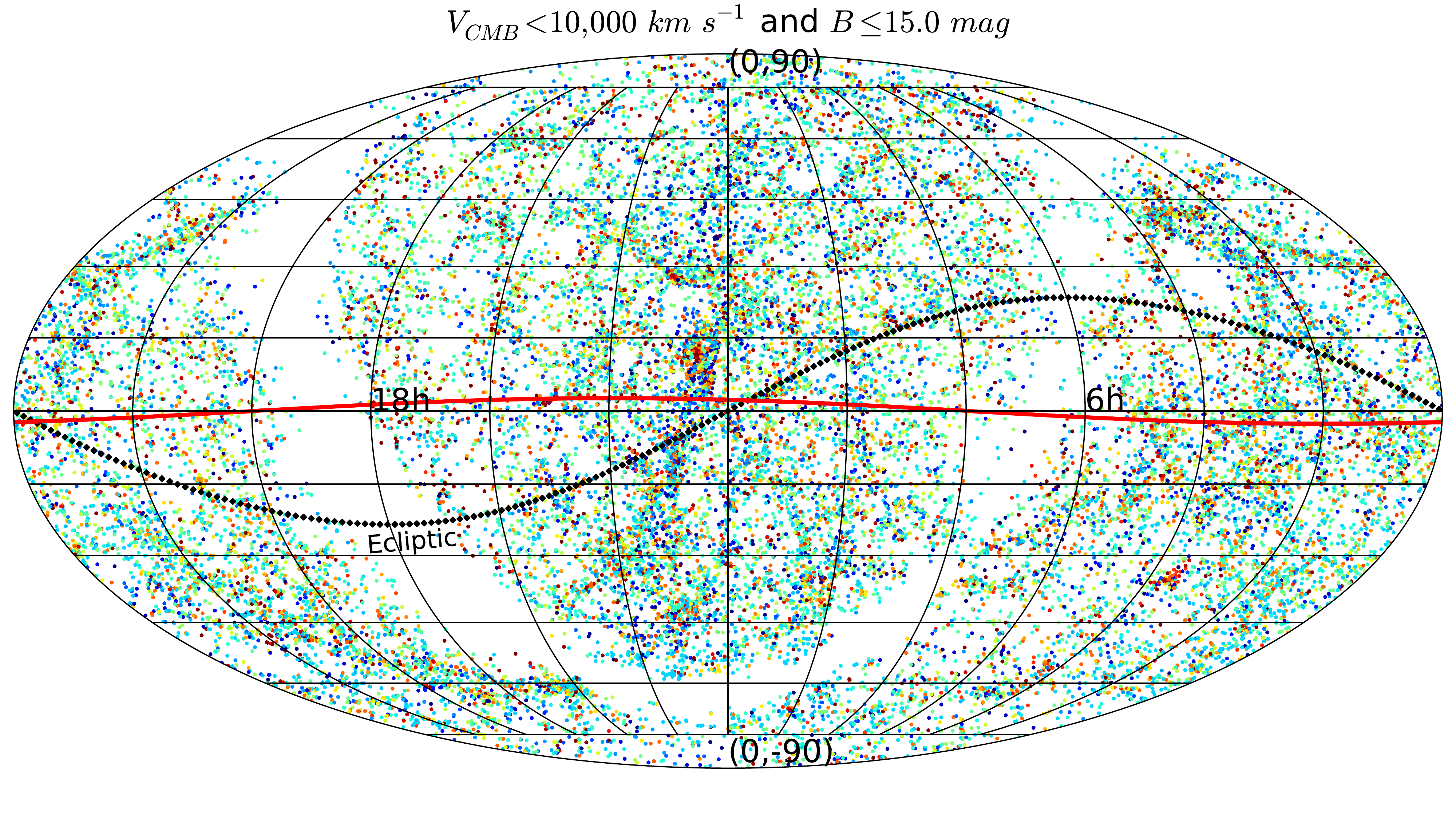}
  \includegraphics[scale=0.22]{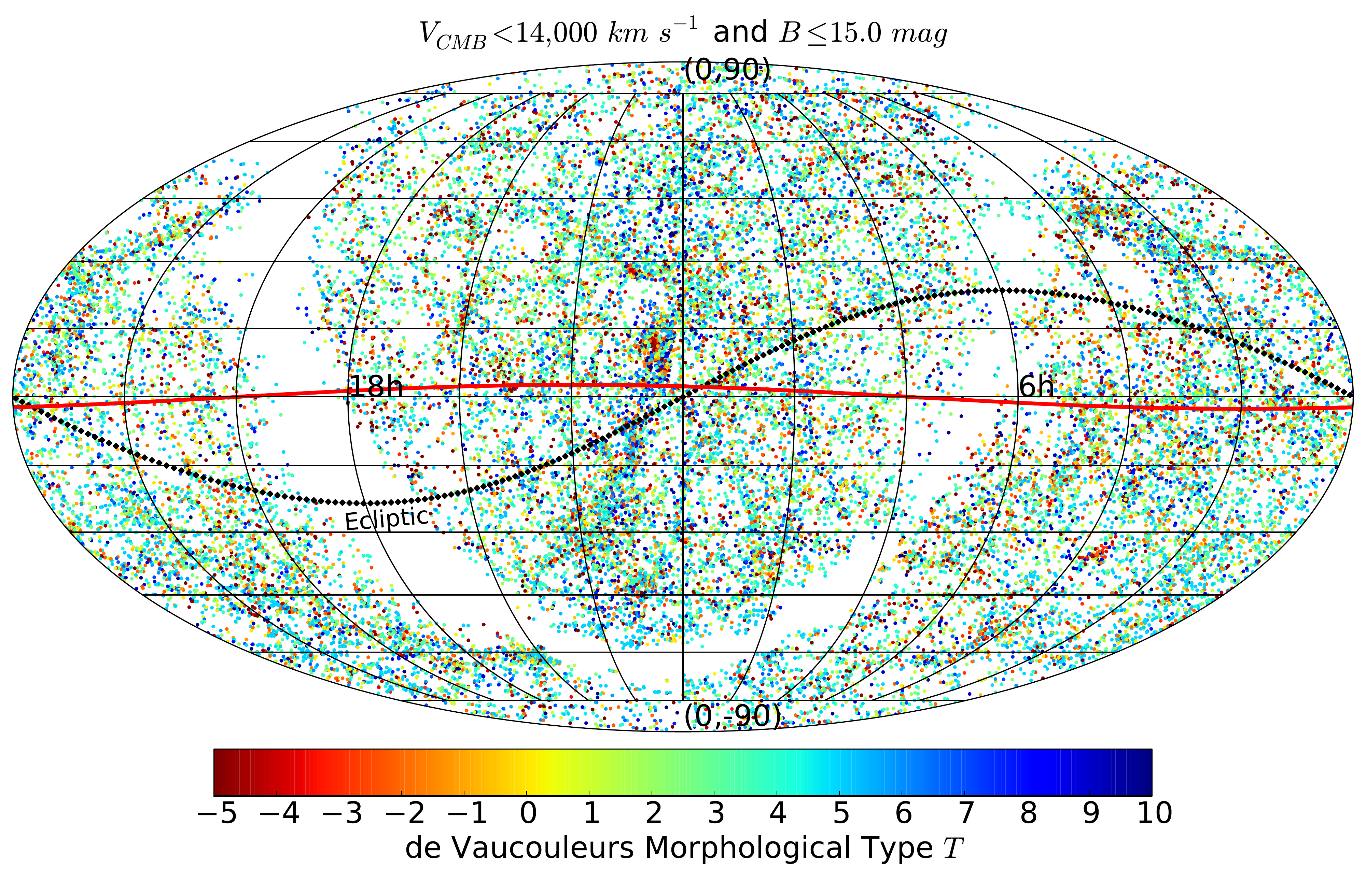}
  \caption{The Equatorial coordinate system distribution of all the galaxies in the HyperLeda database with $B\leq 15.0$ mag and measured $T$ and M$_B$, from top to bottom corresponding to $V_{CMB}<$7,000, 10,000 and 14,000 km/s, respectively. The color code is $T$, the empty parts are the region around the disk of the Milky Way, the plane of the Ecliptic is shown with a dotted line and the pair of hemispheres with the largest difference in the distribution of morphologies are divided by a red solid line. Though the sharp contrast seen in Figure \ref{fig:sky_dist_t} is not visible any longer, the hemispherical asymmetry obtained from the KS test is even more aligned with the Celestial Equator for galaxies with  $V_{CMB}<$ 10,000 and 14,000 km/s. \label{fig:sky_dist_t_maglim}}
  
 \end{center}
\end{figure}

Figure \ref{fig:num_and_dif} shows the number distribution of different $T$ values in the hemisphere pairs corresponding to $D_{max}$ for the magnitude limited samples. We also show the difference in the number of each $T$ bin between the two hemispheres. The largest difference is for the galaxies with $-5\leq T \leq -4$, $-3\leq T \leq -2$, $3\leq T \leq 4$, and $5\leq T \leq 6$, in the order of relative difference. Based on these number differences we see that if $\approx30$\% of galaxies with $3\leq T \leq 4$ in the North and $\approx 34$\% of galaxies with $5\leq T \leq 6$ in the South had their morphological types in the range $4\leq T \leq 5$, and at the same time, $\approx35$\% of galaxies with $-5\leq T \leq -4$ in the North and $\approx40$\% of galaxies with $-3\leq T \leq -2$ in the South had their morphological types in the range $-4\leq T \leq -3$, then the anisotropy would be partially alleviated. However even in that case, the number of galaxies with $4\leq T \leq 5$ would be $\approx17$\% more in the South and the number of galaxies with $-4\leq T \leq -3$ would be $\approx25$\% more in the North. Actually, it is not straightforward to tell what sort of shift in the $T$ values will remove the asymmetry. Figure \ref{fig:sky_dist_t_maglim} shows the distribution of the morphological types of the magnitude limited samples in the Equatorial coordinate system. By comparing this with Figure \ref{fig:sky_dist_t} we can see that the obvious sharp contrast between the north and the south in the case of $V_{CMB}<$10,000 and 14,000 km/s samples is not visible any longer, but the hemispherical asymmetry (shown by the red solid curve) is more aligned with the Celestial Equator. Applying the $B\leq 15.0$ mag limit reduces the level of anisotropy, but does not clean it away.\\

However, the fact that the sky coverage of the distance limited sample is pretty much uniform and it covers a wide range of apparent magnitude as well as a full range of distances, plus the fact that the pair of hemispheres with the largest difference have a similar number of galaxies, raise the question why the distance limited sample, though incomplete, should be biased by a certain type of galaxy in one hemisphere and by a different type in the opposite one.

\begin{table*}
 \begin{center}
  \caption{The results for the samples with  B$\leq$15 mag. With this magnitude limit, around 99\% of the galaxies in the HyperLeda database within all the three radial velocity ranges in our study have a $T$ value. Note that for $V_{CMB}<$10,000 and 14,000 km/s samples, the direction of the $\hat{r}_{max}$ is only $2^{\circ}.6$ away from the Celestial North Pole. \label{t:completeness}}
  
  \begin{tabular}{lcccccccc}
  $V_{CMB}$ $(\rm{km/s})<$  & $\hat{r}_{max}=(l,b)$ & $\alpha_{CNP}$ & $\alpha_{ENP}$ & $D_{max}$ & $p(D_{max})$& $D_{min}$ & $p(D_{min})$\\
  \hline
  7,000 & (112.5,35.7) & 12$^{\circ}$.3 & 14$^{\circ}$.7 & 0.071 &  $1.5\times 10^{-22}$ & 0.016 & 0.15 \\
  10,000 & (123.7,24.6) & $2^{\circ}.6$ & 24$^{\circ}$.8   & 0.069 &  $2.2\times 10^{-31}$& 0.010 & 0.48  \\
  14,000 & (123.7, 24.6) & $2^{\circ}.6$ & 24$^{\circ}$.8   &0.074  & $4.7\times 10^{-43}$& 0.009 & 0.44 \\
  \hline
  \end{tabular}
 \end{center}
\end{table*}

\section{Summary and conclusion}\label{sec:conc}
We presented the first probe of isotropy of the distribution of morphological types of galaxies in the Local Universe. Using the de Vaucouleurs morphological types of more than 60,000 galaxies with radial velocity $V_{CMB}<$14,000 km/s (corresponding to within a distance of $\approx$200 Mpc) from the HyperLeda database, we searched for any directional difference in the distribution of morphological types. We used a hemispherical comparison method and by dividing the sky into two opposite hemisphere pairs, compared the frequency of morphological types, $T$, using a Kolmogorov-Smirnov test. The KS test was applied to hemisphere pairs with the axis of symmetry pointing at the centers of the pixels of a HEALPix grid. This gave us all-sky maps of the KS statistics, $D$. We performed this analysis for three radial velocity ranges, i.e. for galaxies with $V_{CMB}<$7,000, 10,000 and 14,000 km/s (corresponding to distances of about 100, 150 and 200 Mpc).\\ 

The directions $\hat{r}_{max}$ corresponding to the hemisphere pairs with the largest difference, $D_{max}$, were found to be similar for the three distance ranges. These directions are very far from the directions with the largest difference in the number of galaxies. Under the assumption that the galaxy morphologies should be statistically isotropic in the distance ranges under consideration, the probability of obtaining the observed difference or larger from the KS distribution, $p(D_{max})$, was obtained to be $\leq 10^{-133}$. In addition, using a Monte Carlo analysis and by creating 1000 isotropic realizations, the number of realizations with equal or larger $D_{max}$ was found to be zero and the largest values of $D_{max}$ obtained from the 1000 realizations were found to be an order of magnitude smaller than the observed values.\\

Interestingly, the hemispherical asymmetry that we found in the distribution of the morphological types of galaxies is aligned with both the Ecliptic and the Celestial Equator planes. The direction $\hat{r}_{max}$ for the sample with $V_{CMB}<$7,000 km/s is only $12^{\circ}.3$ and $14^{\circ}.7$ away from the Celestial and the Ecliptic North poles, respectively. Using our Monte Carlo analysis with 1000 isotropic realizations, we quantified the significance of the alignment with the Ecliptic to be at the 97.6\% confidence level, and that of the alignment with the Celestial Equator to be at the 97.1\% (both regardless of the extreme significance obtained from the KS test). It may be interesting for the reader to note that the hemispherical asymmetry in the CMB power spectrum discovered in the \textit{WMAP} data and confirmed by the \textit{Planck} satellite is also aligned with the plane of the Ecliptic \citep{planck}.\\

For the other two samples with $V_{CMB}<$10,000 and 14,000 km/s, the observed anisotropy is aligned with the Celestial Equator at the 99.8\% confidence level, with an angular separation of only $5^{\circ}.1$. In general, when looking at the sky distribution of morphological types of the whole sample ($V_{CMB}<14,000$ km/s), the northern sky is more populated by late type galaxies whereas early type galaxies are the dominant type in the southern sky. In particular, the largest difference in the abundance of morphological types is observed to be related to the galaxies with $9\leq T \leq 10$ (i.e. Sm and Im types)  whose number is twice larger towards the Northern sky, and the galaxies with $-1\leq T < 0$ (i.e. S0$^+$ types) whose number is more than twice larger in the Southern sky. Excluding galaxies with large uncertainty on $T$ (i.e. $\sigma_T \geq 3.0$) does not affect the direction of the asymmetry. Also, repeating the analysis on the independent subsamples of galaxies with 7,000$<V_{CMB}$ (km/s)$<$10,000 and 10,000$<V_{CMB}$ (km/s)$<$14,000, yields similar direction for the largest difference in the distribution of $T$.\\

To increase the completeness of our sample, we applied a conservative magnitude limit and included only the galaxies with $B\leq 15.0$ mag in our analysis. This resulted in a decrease in the values of $D_{max}$, though still with a small probability of consistency with the null hypothesis of isotropy of $p(D_{max})\leq 10^{-21}$ for all the three distance ranges. For the sample with $V_{CMB}<$7,000 km/s, the direction of $\hat{r}_{max}$ remained unchanged with respect to the sample without the magnitude limit, however, for the samples with $V_{CMB}<$10,000 and 14,000 km/s and $B\leq 15.0$ mag the direction of $\hat{r}_{max}$ was found to be closer to the Celestial North Pole with only $2^{\circ}.6$ angular separation, i.e. the hemispherical asymmetry is even more aligned to the Celestial Equator than in the samples with the same distance ranges but without the magnitude limit.\\

Using a separate 1000 Monte Carlo realizations in each of which the $T$ values are  shifted randomly within their uncertainties, we quantified the effect of random errors on the anisotropy in the distance limited samples to be negligible. However, for the $V_{CMB}<$7,000 km/s and $B\leq 15.0$ mag sample, shuffling $T$ within $\sigma_T$ can change the direction of anisotropy towards various directions in each realization, meaning that the anisotropy in this sample is not significant. On the other hand, in 97.2\% and 99.9\% of the realizations for $V_{CMB}<$10,000 and 14,000 km/s and $B\leq 15.0$ mag, the direction $\hat{r}_{max}$ is within $32^{\circ}.0$ of the observed direction showing that the anisotropy in these samples (especially in the latter one) is robust against the effect of random errors. For these two sample, the number of galaxies with $-5\leq T < -4$ and with $3\leq T < 4$ is respectively around 50\% and 40\% larger, and the number of galaxies with $-3\leq T < -2$ and $5\leq T < 6$ is respectively around 40\% and 34\% smaller, in the Northern sky. A simple shift (e.g. +1 or -1) in the $T$ values of one hemisphere does not decrease the value of $D_{max}$ (it actually increases it in most cases). Various combinations of shifts in $T$ around $-5\leq T < -2$ and around $3\leq T < 6$ can only partially alleviate the tension in the number of morphological types between north and south.\\

If this significant deviation from isotropy is real and not due to issues with the catalog and the classifications, it could mean that the galaxies in these two opposite directions have had different evolution and/or formation history which would be a major challenge for the Cosmological Principle. However, the fact that the asymmetry has an Equatorial North-South alignment indicates that most probably its source is a systematic bias in the classification of morphological types between the data of the northern and the southern sky possibly due to the fact that different telescopes (with different systematics) were used for observing these galaxies. \\

The morphological types cataloged in RC3 and HyperLeda database have been used for many studies and also some other cataloges and online databases extract the morphological information from them. Our results indicate that some of the studies based on this morphological information may need to be reconsidered. In addition, we should not forget that these classifications have been mostly done visually and it is necessary to put even more effort on making well-defined quantitative and automated approaches for morphological classifications. Further investigations though are indispensable to uncover the exact reason for the anisotropy we found in this study.

\begin{acknowledgements}
We thank the referee for their critical and constructive comments. We are very grateful to Dmitry Makarov for his kind and helpful correspondence and very useful information on the HyperLeda database. We thank Christian Henkel and David Martinez-Delgado for useful discussions, Jan Pflamm-Altenburg, Cristiano Porciani and Hosein Haghi for their constructive comments, Jens Erler for his kind help with HEALPix, and Michael Marks and Patrick Simon for useful discussions on statistics. We acknowledge the usage of the HyperLeda database (http://leda.univ-lyon1.fr), HEALPix package, and the matplotlib plotting library.
\end{acknowledgements}

\bibliography{biblio}

\end{document}